\documentclass[twocolumn]{aastex63}

\usepackage[caption=false]{subfig}
\usepackage{csquotes}

\received{2021 March 23}
\revised{2021 August 13}
\accepted{2021 August 31}
\shorttitle{Population Trends in \textit{Spitzer} Emission of Hot Jupiters}
\shortauthors{Goyal et al.}
\graphicspath{{./}{figures/}}

\begin{document}


\title{Why is it So Hot in Here? Exploring Population Trends in \textit{Spitzer} Thermal Emission Observations of Hot Jupiters using Planet-Specific Self-Consistent Atmospheric Models.}

\correspondingauthor{Jayesh Goyal}
\email{jgoyal@astro.cornell.edu}

\author[0000-0002-8515-7204]{Jayesh M Goyal}
\affiliation{Department of Astronomy and Carl Sagan Institute, Cornell University \\
122 Sciences Drive, Ithaca, NY, 14853, USA}
\affiliation{National Institute of Science Education and Research (NISER), HBNI, \\
Jatni, Khurda-752050, Odisha, India}

\author[0000-0002-8507-1304]{Nikole K Lewis}
\affiliation{Department of Astronomy and Carl Sagan Institute, Cornell University \\
122 Sciences Drive, Ithaca, NY, 14853, USA}

\author[0000-0003-4328-3867]{Hannah R Wakeford}
\affiliation{School of Physics, University of Bristol, \\
HH Wills Physics Laboratory, Tyndall Avenue, Bristol BS8 1TL, UK}

\author[0000-0003-4816-3469]{Ryan J MacDonald}
\affiliation{Department of Astronomy and Carl Sagan Institute, Cornell University \\
122 Sciences Drive, Ithaca, NY, 14853, USA}

\author[0000-0001-6707-4563]{Nathan J Mayne}
\affiliation{Astrophysics Group, School of Physics and Astronomy, University of Exeter, \\ Exeter EX4 4QL, UK}










\begin{abstract}

Thermal emission has now been observed from many dozens of exoplanet atmospheres, opening the gateway to population-level characterization. Here, we provide theoretical explanations for observed trends in \textit{Spitzer} IRAC channel 1 (3.6\,$\micron$) and channel 2 (4.5\,$\micron$) photometric eclipse depths (EDs) across a population of 34 hot Jupiters. We apply planet-specific, self-consistent atmospheric models, spanning a range of recirculation factors, metallicities, and C/O ratios, to probe the information content of \textit{Spitzer} secondary eclipse observations across the hot-Jupiter population. We show that most hot Jupiters are inconsistent with blackbodies from \textit{Spitzer} observations alone. We demonstrate that the majority of hot Jupiters are consistent with low energy redistribution between the dayside and nightside (hotter dayside than expected with efficient recirculation). We also see that high equilibrium temperature planets ($T_{\textup{eq}}$ $\ge$ 1800\,K) favor inefficient recirculation in comparison to the low temperature planets. Our planet-specific models do not reveal any definitive population trends in metallicity and C/O ratio with current data precision, but more than 59 \% of our sample size is consistent with the C/O ratio $\leq$ 1 and 35 \% are consistent with whole range (0.35 $\leq$ C/O $\leq$ 1.5). We also find that for most of the planets in our sample, 3.6 and 4.5 $\micron$ model EDs lie within $\pm$1 $\sigma$ of the observed EDs. Intriguingly, few hot Jupiters exhibit greater thermal emission than predicted by the hottest atmospheric models (lowest recirculation) in our grid. Future spectroscopic observations of thermal emission from hot Jupiters with the James Webb Space Telescope will be necessary to robustly identify population trends in chemical compositions with its increased spectral resolution, range and data precision. 

\end{abstract}


\section{Introduction} \label{sec:intro}

\textit{Spitzer}  observations of exoplanet thermal emission have proven fundamental to population studies of exoplanet atmospheres. To date, \textit{Spitzer} \citep{Werner2004} has been the only space-based observatory available for investigating exoplanets in close orbits to their host stars at infrared wavelengths longer than 2\,$\micron$. \textit{Spitzer's} Infrared Array Camera (IRAC) \citep{Fazio2004} photometric channels 1, 2, 3 and 4 -- centred at 3.6, 4.5, 5.8 and 8\,$\micron$, respectively -- provided early insights into the atmospheres of close-in exoplanets \citep[e.g.][]{Charbonneau2005, Knutson2007,Beichman2018}. However, the cryogen exhaustion in May 2009 rendered IRAC channels 3 and 4 inoperative. Up until the end of the \textit{Spitzer} mission in 2020, IRAC channels 1 (3.6\,$\micron$) and 2 (4.5\,$\micron$) were used extensively for exoplanet detection and atmospheric characterization. 

A number of previous studies have computed/used and investigated thermal eclipse measurements for different populations of hot-Jupiter exoplanets \citep[e.g.][]{Cowan2011, Triaud2014, Schwartz2015, Garhart2020, Melville2020, Baxter2020}. \citet{Garhart2020} and \citet{Baxter2020} (hereafter G20 and B20, respectively) presented \textit{Spitzer} IRAC channel 1 and 2 eclipse depth (ED) measurements for a population of planets. Both of these studies explored potential trends in hot-Jupiter dayside thermal emission as a function of the equilibrium temperature and the assumed atmospheric composition. G20 concluded that the ratio of 4.5 and 3.6 $\micron$ brightness temperatures (T$_{\textup{b}}$), calculated using their EDs, increases with equilibrium temperature, indicating deviation from a planet that emits like a blackbody as found by \citet{Triaud2014}. They also compared the observations presented in their paper with generic track models from \citet{Fortney2005} and \citet{Burrows1997}, instead of carrying out model simulations for each individual planet across a parameter space. B20 also find a similar deviation from blackbody emission spectra. Moreover, B20 also detect a statistically significant transition between hot-Jupiter emission observations above and below 1660 $\pm$ 100\,K, which they suggest as a natural boundary between `hot' and `ultra-hot' Jupiters.  Again without carrying out model simulations for each individual planet, B20, using their generic grid of 1D radiative-convective models spanning a range of C/O ratio (C/O = 0.1, 0.54, 0.84) and metallicity ([M/H] = -1, 0, 1, 1.5) conclude that hot Jupiters statistically favor low C/O ratios (C/O $\leq$ 0.54), where $\sim$0.54 is the value of the solar C/O ratio. \citet{Cowan2011} in their analysis that included both thermal (\textit{J}, \textit{H} , \textit{K} and \textit{Spitzer} bands) as well as optical measurements, concluded that planets with no circulation and no albedo limit temperature of greater than 2400\,K have uniformly low redistribution efficiencies and albedo. A subsequent study by \citet{Schwartz2015} conclusively showed that planets with high irradiation temperatures have low-heat transport efficiency.

Our investigation in this work builds on the previous works, where we apply a planet-specific self-consistent grid of hot Jupiter dayside emission spectra to explain the \textit{Spitzer} IRAC channel 1 and 2 observations from \citet{Garhart2020} and \citet{Baxter2020}. Self-consistent planet-specific model simulations have an advantage of estimating a more accurate $P$-$T$ structure of a given planetary atmosphere compared to generic models. This enhanced accuracy arises from accounting for the host stellar energy deposited in the specific planet's atmosphere, alongside conservation of the energy in the atmosphere itself, when computing a self-consistent $P$-$T$ profile. Consequently, the model emission spectra used to interpret observations, such as those in \textit{Spitzer} IRAC channels 1 and 2, can attain higher accuracy. In the previous works \citep[for e.g.][]{Garhart2020, Baxter2020} self-consistent generic model simulations were used without focusing on model simulations for specific planets (i.e without using specific stellar/planetary parameters for model input). Whereas, in this work, we use self-consistent model simulations performed for the specific parameters of the target planets, spanning a wider parameter space. We use our planet-specific grid to explore population trends across the hot-Jupiter collective. The G20 and B20 studies show a few inconsistencies in their ED and T$_{\textup{b}}$ (and its uncertainty) values for same planets, due to differences in their data reduction techniques and methodology/assumption while computing them. Therefore, in this study we also revisit and compare these datasets.

We delve into the theoretical analysis of \textit{Spitzer} IRAC ED and brightness temperature variations for each planet, thereby investigating any trends in them and any variation of these trends with planetary equilibrium temperature (T$_{\textup{eq}}$). We apply a grid of 1D self-consistent planet-specific models across a wide parameter space, considering variations in their recirculation factor, metallicity, and C/O ratio \citep[see][for further details]{Goyal2020}. Our models incorporate horizontal advection by reducing the incoming flux in each 1D column of the atmosphere by a factor called the \enquote{recirculation factor} \citep{Fortney2007,Goyal2020}, hereafter termed f$_{\textup{c}}$ (see Section \ref{sec:grid_details} for more details). Previous studies did not consider the full range of the variation of f$_{\textup{c}}$ while computing self-consistent Pressure-Temperature ($P$-$T$) profiles and the corresponding emission spectra, which we investigate in this work. It is important to note that f$_{\textup{c}}$ alters the $P$-$T$ profile drastically, therefore is not just a proxy for temperature but also the pressure level probed using the emission spectra. We compare our theoretical EDs across the parameter space with observed EDs from G20 and B20. We find the best-fit parameters as well as those within one sigma of the best-fit model for each planet, with the motivation to identify trends and constraints in f$_{\textup{c}}$, metallicity and the C/O ratio for the population of planets. We also show the detailed analysis for a few planets for which our models are not able to explain the observations and potential reasons for such anomalies. 

This work is structured as follows. We first detail our theoretical model calculations and grid, alongside techniques to compute model EDs and T$_{\textup{b}}$ (model derived and observed both), in Section \ref{sec:grid_details}. In Section \ref{sec:benchmarking}, we benchmark and compare our model emission spectra with those presented in G20. In Section \ref{subsec:results2} we compare the observed EDs and derived T$_{\textup{b}}$ from G20 and B20. The results from our hot-Jupiter thermal emission population analysis are presented in Section \ref{sec:results}, followed by a comparison to the observations of G20 and B20. Finally, we present our conclusions in section \ref{sec:conclusions}.

\section{Techniques and Model details}
\label{sec:grid_details}

In this work, we use a cloud-free planet-specific grid of emission spectra, generated using self-consistent simulated exoplanet atmospheres, building on the work of \citet{Goyal2020} using the \texttt{ATMO}, a 1D-2D radiative-convective equilibrium model for planetary atmospheres \citep{Amundsen2014, Tremblin2015, Tremblin2016, Drummond2016, Goyal2018}. `Self-consistent' here implies that the final computed $P$-$T$ profiles (once they reach convergence) are in radiative-convective equilibrium consistent with equilibrium chemical abundances, under the constraints of hydrostatic equilibrium and conservation of energy in each atmospheric layer and the atmosphere as a whole \citep[see][for more details]{Goyal2020}.  Since the work of \citet{Goyal2020} did not contain model atmospheres for 14 of the exoplanets presented by both G20 and B20, we generated model atmospheres and spectra for these additional planets using exactly the same model and grid parameters as \citet{Goyal2020}. The new planets considered here are: KELT-2b, KELT-3b, Qatar-1b, WASP-14b, WASP-18b, WASP-36b, WASP-46b, WASP-64b, WASP-65b, WASP-75b, WASP-77b, WASP-87b, WASP-100b, and WASP-104b. Observations of KELT-7 are present in G20 but not in B20. We note that while our analysis covers all the planets with observations presented by G20, it does not extend to all the planets considered by B20.

f$_{\textup{c}}$ parameterizes the redistribution of input stellar energy in the planetary atmosphere, by the dynamics, where a value of $1$ equates to no redistribution (hottest dayside temperature), while $0.5$ represents efficient redistribution. The value of 0.5 f$_{\textup{c}}$ indicates 50\% of the total incoming stellar energy is advected to the nightside (the side of the planet facing away from the star), while 0.25 f$_{\textup{c}}$ indicates that 75\% of the total incoming stellar energy is advected to the night side. Given the short orbital periods, we assume all the hot Jupiters in this study are tidally locked and therefore that the f$_{\textup{c}}$ is a good description of the heat transport in the atmosphere. The model atmospheres and emission spectra for each planet are computed at four different recirculation factors (f$_{\textup{c}}$ = 0.25, 0.5, 0.75, 1.0), six metallicities (0.1, 1, 10, 50, 100, 200; all in units of $\times$\,solar) and six C/O ratios (0.35, 0.55, 0.7, 0.75, 1.0, 1.5), giving a total of 144 simulated model spectra per planet. For most planets, our 1D model atmospheres were generated spanning the full range of the parameter space. However, as noted in \citet{Goyal2020}, in 5\%-10\% of the cases the combination of parameters for some of the planets failed to produce a realistic thermal structure (convergence not reached). We do not consider those models in this study.

We consider a wide range of opacity sources relevant to hot giant planets in our models. All model atmospheres include line opacity due to H$_{2}$O, CO$_2$, CO, CH$_4$, NH$_3$, Na, K, Li, Rb, Cs, TiO, VO, FeH, CrH, PH$_3$, HCN, C$_{2}$H$_{2}$, H$_{2}$S, SO$_{2}$, H$^-$, and Fe. The abundances of these species are determined via the assumption of thermochemical equilibrium, consistent with the radiative-convective equilibrium $P$-$T$ profile. Besides these molecular, atomic, and ionic opacities, we also include collision-induced absorption (CIA) due to H$_2$-H$_2$ and H$_2$-He. The line-by-line cross-sections (resolution of 0.001 cm$^{-1}$ evenly spaced in wavenumbers) of these opacities are used to generate correlated-k tables as a function of wavenumber, temperature, and pressure for each gaseous species. We use these correlated-k tables to generate $P$-$T$ profiles and emission spectra for each model atmosphere, given the composition, deploying the random overlap technique \citep{Amundsen2017} to combine k-tables of different gaseous species \citep[see][for more details]{Goyal2018,Goyal2020}.

For the stellar spectra incident on each planet, we use models from the Phoenix BT-Settl\footnote{\url{https://phoenix.ens-lyon.fr/Grids/BT-Settl/AGSS2009/SPECTRA/}} grid \citep{Allard2012, Rajpurohit2013}. For each planet, we choose the stellar model closest to the host star's observed temperature, gravity, and metallicity as listed in the TEPCat\footnote{\url{http://www.astro.keele.ac.uk/jkt/tepcat/allplanets-ascii.txt}} database \citep{Tepcat2011}. We also adopt all the other parameters required for model initialization (stellar radius, planetary radius, planetary equilibrium temperature, planetary surface gravity, and semi-major axis) from TEPCat -- as shown in Table S3 of the online supplementary material in \citet{Goyal2020} and in Table 2 in Appendix \ref{app:planet_data} for the remaining 14 planets, for which the grid of model simulations have been developed in this work. 

\subsection{Model Eclipse Depth and Brightness Temperature Calculations}
\label{sec:bt}

To compare ED observations in \textit{Spitzer} IRAC channels 1 and 2 with simulated model emission spectra, we need to compute model EDs in these channels. ED is the product of planet-to-star flux ratio and the transit depth $(R_p/R_s)^2$ of the planet. We compute planet-to-star flux ratio $\Big($$\frac{F_{\rm int(p)}}{F_{\rm int(s)}}$$\Big)$ in \textit{Spitzer} IRAC channels using 

\begin{equation} 
     F_{\rm int(p)} = \frac{\int_{0}^{\infty} F_{\rm p}(\lambda) \lambda R(\lambda)d\lambda}{\int_{0}^{\infty} \lambda R(\lambda)d\lambda} 
     \label{eq:BT_model21}
\end{equation}

and

\begin{equation} 
     F_{\rm int(s)} = \frac{\int_{0}^{\infty} F_{\rm s}(\lambda) \lambda R(\lambda)d\lambda}{\int_{0}^{\infty} \lambda R(\lambda)d\lambda}, 
     \label{eq:BT_model22}
\end{equation}

where $F_{\rm p}(\lambda)$ and $F_{\rm s}(\lambda)$ are the wavelength $\lambda$ dependent planet and stellar fluxes from the models, respectively. These model planet and stellar flux are integrated in Equation \ref{eq:BT_model21} and \ref{eq:BT_model22} by convolving with the response function $R(\lambda)$ of the given channel to obtain 
integrated planet (F$_{\rm int(p)}$) and stellar (F$_{\rm int(s)}$) flux. The transit depth obtained from the TEPcat (same database as our model input parameters) database when multiplied to this planet-to-star flux ratio in the given channel, gives us the final model ED in that channel.  We note that we employed EDs from G20 and not B20 while fitting to observations in Section \ref{subsec:results3}, because G20 used an additional dilution correction in their data reduction to account for a stellar companion (see Section \ref{subsec:results2} for more details).

Brightness temperature (T$_{\textup{b}}$) is a quantity commonly used in remote sensing, to quantify the blackbody temperature of the remotely sensed object. However, there is a subtle difference between the two approaches used to compute T$_{\textup{b}}$ for exoplanets. One approach is to compute \emph{theoretical} T$_{\textup{b}}$ using model emission spectra. Alternatively, one may convert the observed EDs into an equivalent T$_{\textup{b}}$. To highlight these differences, we demonstrate this computation using both model simulated emission spectra and observed EDs.

For this work we need to compute T$_{\textup{b}}$ in the \textit{Spitzer} IRAC channel 1 centered at wavelength ($\lambda$) of 3.6\,$\micron$ and channel 2 centered at 4.5\,$\micron$.

We compute model T$_{\textup{b}}$ $\big($T$_{\textup{b(model)}}$$\big)$ via a two-step procedure. First, we compute the integrated model planet flux in the required channel using equation \ref{eq:BT_model21}. 

Secondly, we invert the Planck function to obtain the temperature with equivalent planet flux in the given channel using

\begin{equation} 
     T_{\textup{b(model)}} = \frac{hc}{\lambda k_b}\Big[{\ln\Big(\frac{2hc^2}{\lambda^5\frac{F_{\rm int(p)}}{\pi}} + 1\Big)}\Big]^{-1},
     \label{eq:BT_model1}
\end{equation}
where $h$, $c$, and $k_b$ are Planck's constant, the speed of light, and Boltzmann's constant, respectively. We note that division by the factor of $\pi$ converts the integrated planet flux ($F_{\rm int(p)}$) to specific intensity, required by the inverse Planck equation. 

Alternatively, we derive observed T$_{\textup{b}}$ $\big($T$_{\textup{b(obs)}}$$\big)$, using the EDs from G20, following a similar methodology as described by B20 using

\begin{equation} 
     T_{\textup{b(obs)}} = \frac{hc}{\lambda k_b}\Big[{\ln\Big(\frac{2hc^2}{\lambda^5\frac{F_{\rm int(s)}}{\pi}\frac{ED}{TD}} + 1\Big)}\Big]^{-1},
     \label{eq:BT_model3}
\end{equation}
where $F_{\rm int(s)}$ is the integrated stellar flux (computed using equation \ref{eq:BT_model22}), $ED$ is the eclipse depth in the given channel from G20, and $TD$ is the white light transit depth $(R_p/R_s)^2$. Since the $TD$ values are not quoted in G20 or B20, we use the transit depth values from ExoMAST\footnote{\url{https://exo.mast.stsci.edu/}}. We note that the choice of database for TD values can lead to substantial variation in derived Tb of the planet. We use BT-Settl model stellar spectra, as described in the previous subsection to compute $F_{\rm int(s)}$. This integrated stellar flux $F_{\rm int(s)}$ and $TD$ is then removed from ED to obtain planetary flux, which is finally used to compute observed T$_{\textup{b}}$ of the planet. While solving equation \ref{eq:BT_model3} we are integrating both, the blackbody planetary and model stellar flux over the \textit{Spitzer} bandpass and fitting them to obtain observed T$_{\textup{b}}$ for a given value of ED, therefore this is iteratively solved to obtain a T$_{\textup{b}}$ value that closely matches the observed ED value. We note that we iterate between $\pm$1000\,K of the obtained T$_{\textup{b}}$ using an inverse Planck function while optimizing for the best-fit ED, as opposed to $\pm$200\,K used in B20. This was required because for some planets the ED fit with $\pm$200\,K was insufficient, due to the lower range of the explored temperature. The uncertainties on the observed T$_{\textup{b}}$ are computed using minimum and maximum value of the EDs propagated through equation \ref{eq:BT_model3} to obtain minimum and maximum T$_{\textup{b}}$. The 1$\sigma$ uncertainty for the best fit T$_{\textup{b}}$ is then taken as the mean of this minimum and maximum T$_{\textup{b}}$, similar to the methodology adopted in B20. Our calculated T$_{\textup{b}}$ from observed ED and their uncertainties for all the planets in this analysis are tabulated in Table 1 in Appendix \ref{app1}. Our uncertainties are very similar to that of B20 but very different from that in G20. To understand the uncertainties obtained by G20 we also computed the uncertainty in T$_{\textup{b}}$ by propagating the error in ED through equation \ref{eq:BT_model3} (error propagation methodology). However, we were unable to obtain the uncertainties on T$_{\textup{b}}$ as in G20 (see Section \ref{subsec:results2} for more details). 

\section{Benchmarking}\label{sec:benchmarking}

\begin{figure}
\centering
\includegraphics[width=1.0\columnwidth]{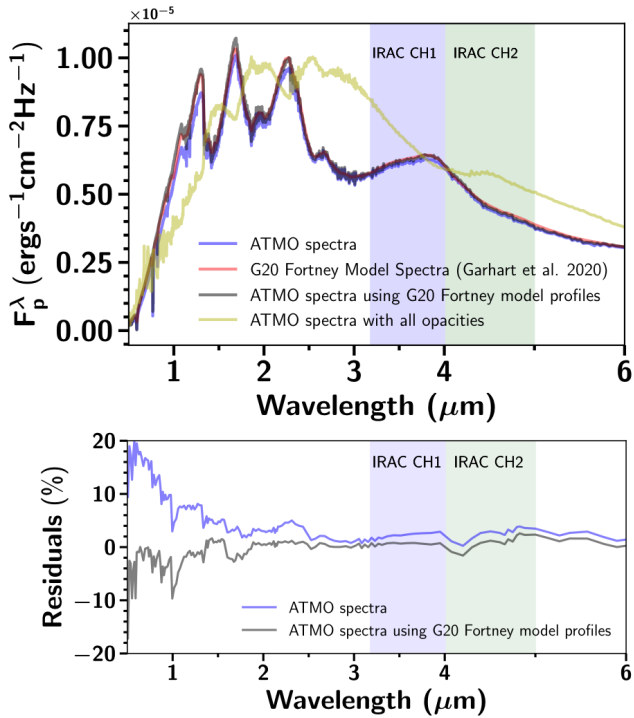}
\caption{Top panel: comparison between our \texttt{ATMO} emission spectrum model and a reference model from G20. Four model spectra are shown: the planetary emission spectra from \texttt{ATMO} (blue) using the exact same opacities as in G20, the ``G20 Fortney model spectra'' (red), an \texttt{ATMO} model using the same $P$-$T$ profile and chemical abundances as the ``G20 Fortney model spectra'' (black) and ATMO model spectra with all opacities (yellow). In all cases, the planet parameters are the same as those described in caption of Figure 19 in G20. \textit{Spitzer}  IRAC channels 1 and 2 are shaded in blue and green, respectively. Bottom panel: flux difference (residuals) in percentage between the ``G20 Fortney model spectra'' and \texttt{ATMO} model emission spectra are shown in the top panel in blue and black. The \texttt{ATMO} model spectra with all opacities being too different from ``G20 Fortney model spectra'' is not shown here.}
\label{fig:benchmark1}
\end{figure}

Before presenting the results of our analysis, we first benchmark our emission spectrum model against G20. While model emission spectra from \texttt{ATMO} have previously been benchmarked against various published model spectra \citep{Baudino2017, Malik2018}, here we compare our \texttt{ATMO} model to Figure 19 of G20 (hereafter, \enquote{G20 Fortney model}), with exact same planet parameters. Figure~\ref{fig:benchmark1} (top panel) shows the planet flux from \texttt{ATMO} and the \enquote{G20 Fortney model spectra}, using their respective $P$-$T$ profiles and chemical abundances. Figure~\ref{fig:benchmark1} (bottom panel) shows residuals (differences) between both the model spectra. The agreement between both the model spectra is quite good, especially in \textit{Spitzer} IRAC channels 1 and 2, where the differences are within +5\%. This  +5\% (rather than $\pm$5\%) difference is because the overall spectral flux in the \enquote{G20 Fortney model spectra} is greater than in our \texttt{ATMO} model spectra, since the $P$-$T$ profile for the \enquote{G20 Fortney model spectra} is $\sim$100-200\,K hotter than the \texttt{ATMO} $P$-$T$ profile between the $\sim$0.1 and 1 bar pressure levels. We note that, for a fair comparison in this benchmarking test, we removed TiO, VO, Li, Rb, Cs, FeH, HCN, SO$_2$ and C$_2$H$_2$ opacities while computing the $P$-$T$ profiles and spectra in \texttt{ATMO} (as these opacities are not included, or these species have very low abundances, in the \enquote{G20 Fortney model}). However, all these opacities are included in our other simulations used throughout this work. Therefore, when using a similar model setup as in G20, the agreement between both models is very good. The differences in $P$-$T$ profiles and chemical abundances drive discrepancies between the model spectra. To illustrate this, Figure~\ref{fig:benchmark1} (top panel) shows that the \texttt{ATMO} model spectra with all opacities is substantially different from the \enquote{G20 Fortney model spectra} -- here, due to the inclusion of TiO/VO in our simulation leading to the formation of a temperature inversion. 

The larger differences in residuals that we see for $\lambda$ $<$ 1\,$\micron$ can be attributed to different model choices: in particular, the species included in the equilibrium chemistry computation, the rainout condensation methodology, line-list sources and the pressure broadening prescription for alkali resonance lines. Optical absorbers such as Na and K are especially affected by these model choices, leading to larger differences for $\lambda$ $<$ 1\,$\micron$. However, we do not focus on this region of the spectrum in this study. Therefore, a detailed benchmarking exercise for optical wavelengths is beyond the scope of this work. In Figure~\ref{fig:benchmark1} (top panel) we also show the \texttt{ATMO} model planet flux derived using input $P$-$T$ profiles and chemical abundances from the \enquote{G20 Fortney model}, resulting in excellent agreement between the two models, especially in the \textit{Spitzer} IRAC channels. The residuals for this model can also be seen in the bottom panel, where the differences are within $\pm$2\% in \textit{Spitzer} IRAC channels 1 and 2. The differences in this case can be directly attributed to differences in line-lists and cross-section computations. The details of line-list sources can be found in \citet{Goyal2020} for the \texttt{ATMO} model and in \citet{Marley2021}, for the \enquote{G20 Fortney model}.

In addition, we benchmarked our model T$_{\textup{b}}$ calculations by using an isothermal emission spectrum with a known temperature as its input. In this case, the accuracy of the T$_{\textup{b}}$ computed from our simulations was within 2\,K of the assumed isothermal temperature.

\section{Observational Comparison between G20 and B20}
\label{subsec:results2}

\begin{figure*}
\centering
\includegraphics[width=\textwidth]{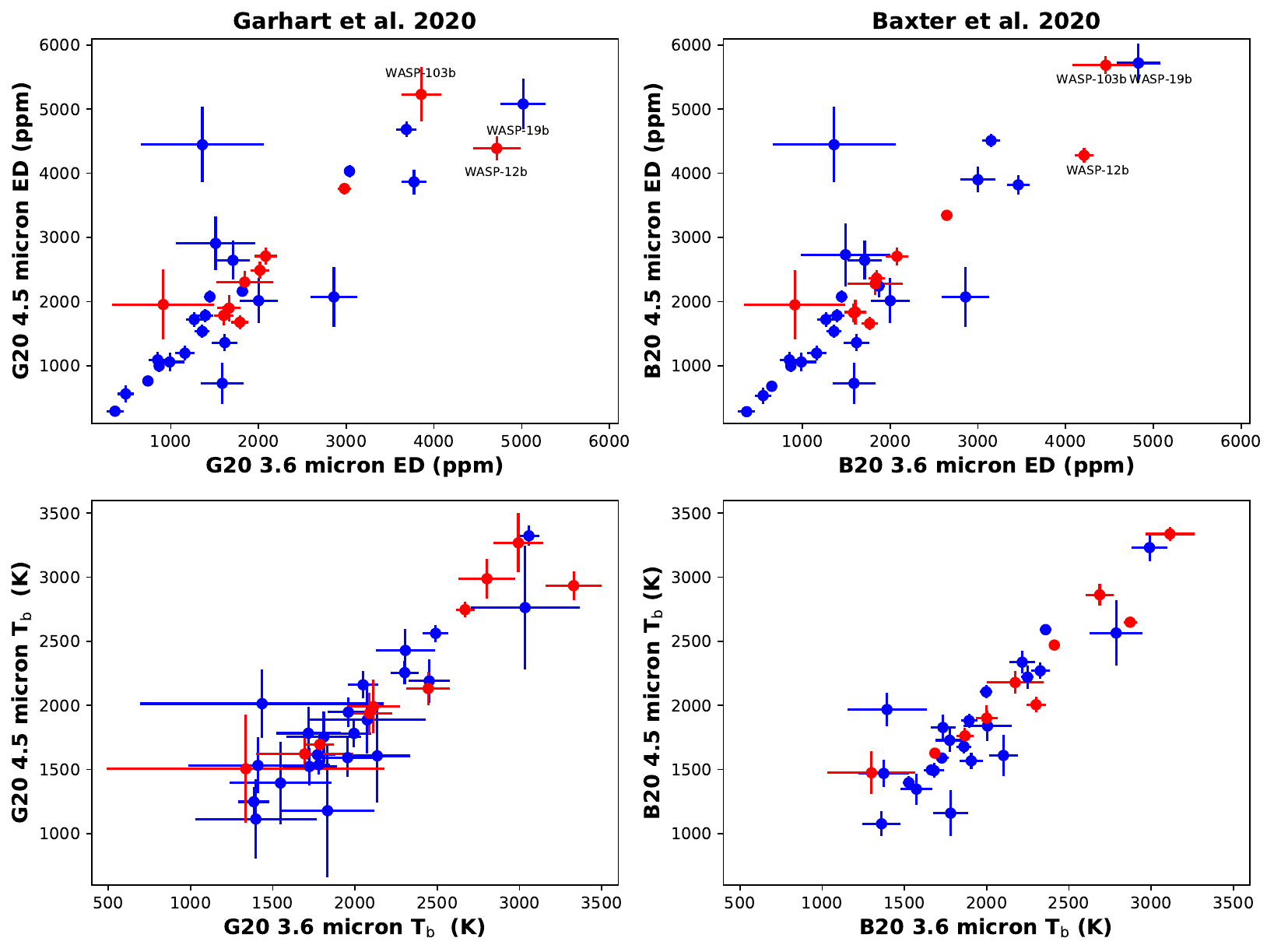}
\caption{Top: observed EDs from G20 and B20 in the \textit{Spitzer} IRAC 3.6 and 4.5\,$\micron$ channels. Planets for which a dilution correction has been applied (as detailed in G20) are shown in red. Planets WASP-12b, WASP-19b and WASP-103b labeled in these plots are examples of planets with quite large differences in ED between G20 and B20. Bottom: same, but comparing the brightness temperature T$_{\textup{b}}$ from G20 and B20. The differences in T$_{\textup{b}}$ are quite large for many of the planets, especially the uncertainties in G20 are substantially larger than B20.}
\label{fig:GB20}
\end{figure*}

Recently B20 and G20 presented independent population-level studies of available \textit{Spitzer} eclipses. Before comparing predictions from our planet-specific self-consistent grid of model simulations to the observed values, it is important to understand the possible inconsistencies in those observed values that might arise due to choices in data reduction techniques and assumptions made in the T$_{\textup{b}}$ calculation. Here we compare the observations and results of G20 and B20. Figure~\ref{fig:GB20} (top panels) compares the EDs from G20 (left) and B20 (right) for all the planets in G20. We see that the agreement is quite good for most planets. However, some of the planets display key differences, e.g., WASP-103b in both the 3.6 and 4.5\,$\micron$ channels ($\sim$ 600 and 450 ppm, respectively), WASP-12b at 3.6\,$\micron$ ($\sim$ 500 ppm), and WASP-19b at 4.5\,$\micron$ ($\sim$ 700 ppm). It is important to note that the EDs for both WASP-103b and WASP-12b from G20 include a dilution correction due to a companion star (see Section 3.5 in G20 for more details), which is not included in B20.

We compare the T$_{\textup{b}}$ from G20 and B20 in the lower panels of Figure~\ref{fig:GB20}. Unlike the EDs, we see that the T$_{\textup{b}}$ display more notable differences. In particular, the T$_{\textup{b}}$ error bars from G20 are much larger than from B20. As shown in Table~\ref{target_table}, and described earlier in Section \ref{sec:bt}, we also calculate our own T$_{\textup{b}}$ from observed EDs, for which the error bars (see Section \ref{sec:bt} for more details) are in close agreement with that of B20. It is unclear why the T$_{\textup{b}}$ error bars from G20 are so large. We used different methodologies to compute uncertainty in T$_{\textup{b}}$ (see Section \ref{sec:bt} for more details), but we could not reproduce uncertainties as high as those obtained by G20. We need to inflate our uncertainties by more than 100\% for many of the planets to match the T$_{\textup{b}}$ uncertainties in G20. The differences in the T$_{\textup{b}}$ values between the different studies are mainly because of the assumed planetary/stellar parameters and the methodology used in the calculation of the T$_{\textup{b}}$ itself. While G20 used ATLAS stellar models, B20 performed tests using a range of stellar models before ultimately adopting PHOENIX models for their final T$_{\textup{b}}$ calculations. We also use PHOENIX models for our calculations, with a similar methodology to compute T$_{\textup{b}}$ as that used in B20.

In summary, we conclude that the differences in EDs from G20 and B20 are negligible, except for some planets with high T$_{\textup{eq}}$ and the systems that require dilution correction. However, the differences in T$_{\textup{b}}$, and especially their uncertainties, are quite substantial. Moreover, the computation of T$_{\textup{b}}$ involves assumptions about the stellar spectra and transit depth (see Equation~\ref{eq:BT_model3}). Therefore, in what follows we present our analysis and comparison between \textit{Spitzer} IRAC observations and our theoretical models in terms of EDs instead of T$_{\textup{b}}$.

\section{Results and Discussion} 
\label{sec:results}

We computed the model T$_{\textup{b}}$ and EDs in \textit{Spitzer} IRAC channels 1 and 2, as described in Section \ref{sec:bt}, for all the model simulations in our grid for a given planet. These model simulations for each planet span a wide range of recirculation factor, metallicities and C/O ratios (see Section \ref{sec:grid_details}). In this section, we first show the theoretical trends in the 3.6 and 4.5\,$\micron$ EDs using examples of certain planets in sub-section \ref{subsec:theory_trends}. The interpretation of the \textit{Spitzer} observations from G20 for the full population of planets is detailed in sub-section \ref{subsec:results3}. Finally, we highlight a few interesting cases where large anomalies exist between the predictions from the model atmospheres and \textit{Spitzer} observations in sub-section \ref{subsec:results4}.

\subsection{Theoretical Trends in Hot Jupiter Emission Spectra}
\label{subsec:theory_trends}

\begin{figure*}
\centering
\includegraphics[width=\textwidth]{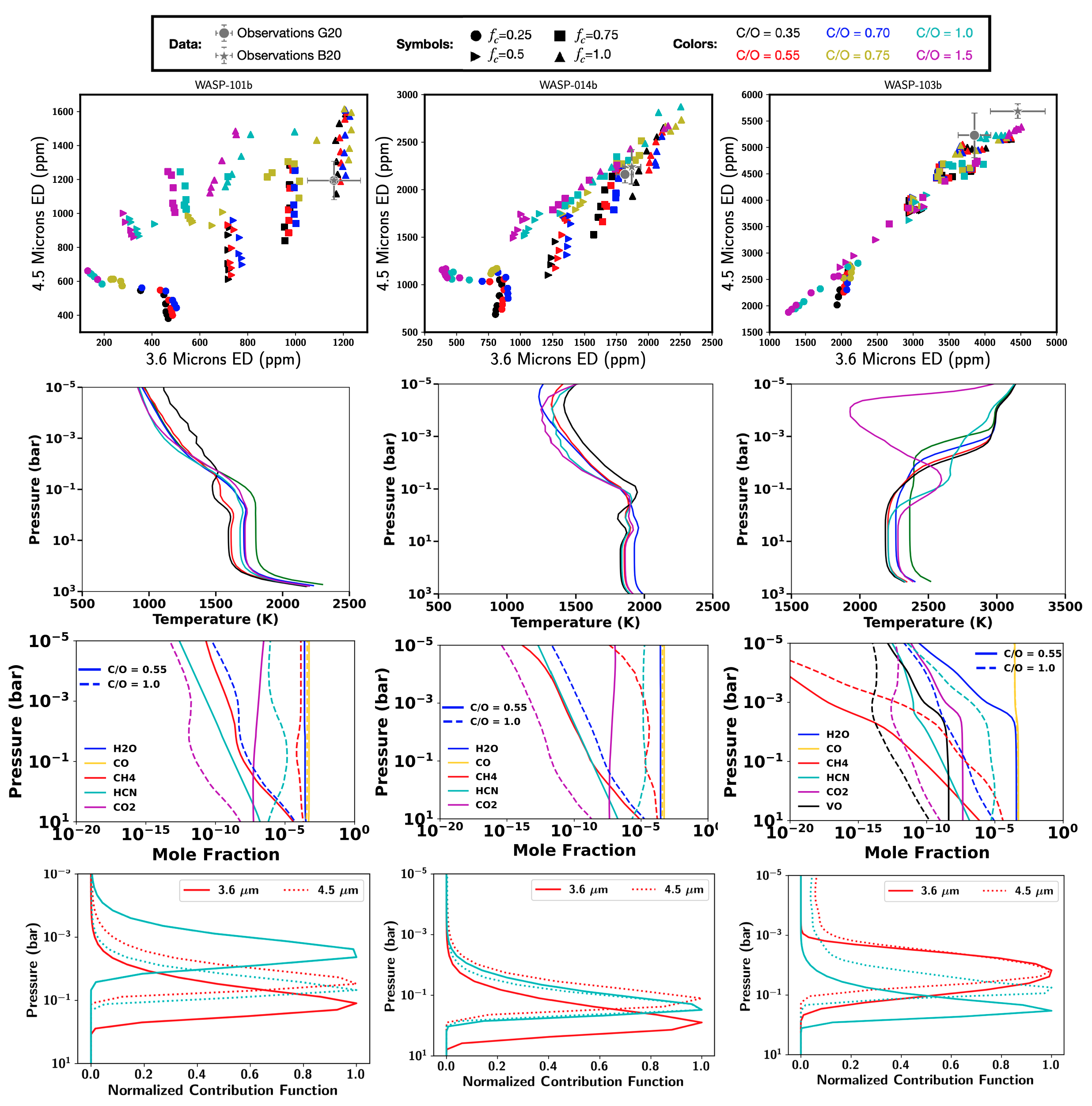}
\caption{First row: model EDs in the 3.6 and 4.5 $\micron$ \textit{Spitzer} bands for WASP-101b, WASP-14b and WASP-103b. The markers correspond to different f$_{\textup{c}}$ values, while the colors correspond to different C/O ratios (see the legend). The observed EDs for each planet from G20 (gray circle with uncertainties) and B20 (gray star with uncertainties) are overlaid for comparison. The observed EDs for WASP-101b from G20 and B20 overlap. Second row: the self-consistent $P$-$T$ profiles for each planet, for $f_{\textup{c}} = 0.5$ and solar metallicity, as C/O varies. Third row: chemical abundances at different atmospheric (pressure) layers for certain important chemical species with strong opacities in either the 3.6 and 4.5 $\micron$ \textit{Spitzer} bands, at C/O ratios of 0.55 (solid line) and 1.0 (dashed line), with the corresponding consistent $P$-$T$ profiles shown in the second row. Fourth row: contribution functions at 3.6 and 4.5 $\micron$ for C/O ratios of 0.55 (red) and 1.0 (cyan), for the $P$-$T$ profiles and chemical abundances shown in the second and third row, respectively.}
\label{fig:theory}
\end{figure*}

Model emission spectra for different planets change as a function of f$_{\textup{c}}$, metallicity and the C/O ratio. This can lead to a range of different values for the \textit{Spitzer} channel 1 and 2 EDs, i.e., their ratios can be strongly influenced by these three atmospheric parameters. In this section, we discuss the variation in these EDs and note some important trends. 

We illustrate these trends for three planets in Figure~\ref{fig:theory}. In the top panels, we show model EDs in \textit{Spitzer} channel 1 (3.6\,$\micron$) and channel 2 (4.5\,$\micron$) for WASP-101b (T$_{\textup{eq}}$=1559\,K), WASP-14b (T$_{\textup{eq}}$=1893\,K), and WASP-103b (T$_{\textup{eq}}$=2513\,K), spanning a range of f$_{\textup{c}}$ values (different markers) and C/O ratios (different colours). The observed EDs from G20 and B20 are overlaid for comparison. In the panels on the second row from the top, we show the corresponding $P$-$T$ profiles for a range of C/O ratios, assuming a solar metallicity and efficient recirculation (f$_{\textup{c}}$ = 0.5). In the panels on the third row, we show chemical abundances at C/O ratio of 0.55 (solar) and 1.0. Finally, in the last row we show the 3.6 and 4.5\,$\micron$ contribution functions for these three planets, at C/O ratios of 0.55 (solar) and 1.0. As an example, we include an ED plot for WASP-101b, similar to that in the top panel in Figure~\ref{fig:theory}, in Appendix \ref{app:metallicity_plot}, but with different colors representing models with different metallicities.

We see a clear structural change in how the 3.6 vs. 4.5\,$\micron$ EDs cluster for different atmospheric parameters between the lowest T$_{\textup{eq}}$ planet (WASP-101b) and the highest temperature planet (WASP-103b). Namely, the spread of points decreases as T$_{\textup{eq}}$ increases, indicating that the change in percentage ED between 3.6 and 4.5\,$\micron$ decreases with increasing T$_{\textup{eq}}$ for the variations in model C/O ratio and metallicity. For the extremely high T$_{\textup{eq}}$ planets, like WASP-103b, the 3.6 and 4.5\,$\micron$ EDs follow a strongly linear correlation across a wide range of our model grid parameters. However, the 4.5\,$\micron$ ED is consistently larger (by $\sim$500 ppm) than the 3.6\,$\micron$ ED for WASP-103b. We also notice that model simulations with the same C/O ratio (same color in Figure~\ref{fig:theory}) have an almost vertical structure, indicating a constant ED in the 3.6\,$\micron$ channel while it varies substantially in the 4.5\,$\micron$ channel. This trend is especially strong for WASP-103b in the vertical branch close to $\sim$ 2000 ppm in the 3.6 $\micron$ channel. This trend results from the change in metallicity affecting primarily the 4.5 $\micron$ channel ED, as explained in detail later in this section for WASP-101b. In contrast, the model simulations with the same metallicity (same color in the left side of Figure~\ref{appfig:metal} shown in Appendix \ref{app:metallicity_plot}) show horizontal structure indicating constant ED in the 4,5\,$\micron$ band. This trend results from the change in C/O ratio primarily effecting the 3.6 $\micron$ channel ED, as explained in detail below for each planet. 

The $P$-$T$ and chemical abundance profiles for the three planets and their contribution functions (all shown in Figure~\ref{fig:theory}) can shed more light into differences between 3.6\,$\micron$ vs. 4.5\,$\micron$ model EDs. For WASP-101b, greater atmospheric depths are probed at 3.6 $\micron$ compared to 4.5\,$\micron$ for a solar C/O ratio, since there is no dominant source of opacity at 3.6 $\micron$ while at 4.5 $\micron$ CO strongly absorbs. The absorption cross-section of CO$_2$ is almost an order of magnitude greater than CO at 4.5 $\micron$ \citep[see Figure 2b in][]{Goyal2020}, but it has a lower abundance (see the chemical abundances in the third row from the top in Figure~\ref{fig:theory}). Therefore, the impact of CO$_2$ on the ED is less than that for CO in this band. At a C/O ratio of 1.0 (cyan color) the trend is reversed, with low pressures being probed at 3.6 $\micron$ and comparatively deeper pressures being probed at 4.5 $\micron$. This leads to lower EDs at 3.6 $\micron$ for a C/O ratio of 1.0 as compared to a solar C/O ratio, as seen in the top panel showing variation in EDs due to the C/O ratio. Fundamentally, this is a result of the change in chemical abundances (see the chemical abundance plots in the third panel from the top in Figure 3), especially the increase in HCN and CH$_4$ at a C/O ratio of 1.0. HCN and CH$_4$ have strong opacity in the 3.6 $\micron$ band, thus the increase in the abundance of these species makes the atmosphere optically thick, therefore allowing only comparatively lower pressure layers with lower temperatures to be probed (see the $P$-$T$ profiles in the middle panel). The change in the atmospheric layers being probed in 4.5 $\micron$ due to change in C/O ratio is very small, as seen in the contribution functions. Moreover, the change in CO abundance with C/O ratio is also small. Therefore, for WASP-101b the change in the 4.5 $\micron$ ED with the change in C/O ratio is smaller compared to the 3.6 $\micron$ ED. In contrast, the change in 4.5 $\micron$ ED with metallicity is larger compared to 3.6 $\micron$ (see Figure~\label{appfig:metal} in Appendix \ref{app:metallicity_plot}). This is driven by the increase in the abundances of CO and CO$_2$ as metallicity increases, making the atmosphere optically thicker at 4.5 $\micron$, thereby probing comparatively low pressures (and hence low temperatures), and finally leading to a lower 4.5 $\micron$ ED at higher metallicities compared to that at solar metallicity. 

For WASP-14b, a similar pattern of change in the layers being probed with change in that C/O ratio (and therefore the EDs) can be seen in the contribution function plots, albeit to a lesser degree. Similar to WASP-101b, at a solar C/O ratio deeper pressures are probed at 3.6 $\micron$ compared to 4.5 $\micron$. However, unlike WASP-101b, at a C/O ratio of 1.0 there is a small change in the pressure levels being probed (towards shallower pressures) at 3.6 $\micron$. The change is such that in both the channels almost same atmospheric layers are probed. This is mainly due to the lower CH$_4$ abundance in the upper atmosphere, as seen in the chemical abundance plots, thereby allowing deeper layers to be probed at 3.6 $\micron$ (compared to WASP-101b). Therefore, the percentage differences in EDs between both the channels due to the change in C/O ratio is smaller for a recirculation factor of 0.5 compared to WASP-101b. However, at lower recirculation factors (cooler $P$-$T$ profiles) the differences in EDs are similar to that of WASP-101b. This decrease in the differences between 3.6 $\micron$ and 4.5 $\micron$ channel model EDs due to the change in C/O ratio, with an increase in the recirculation factor (hotter $P$-$T$ profiles), highlights the temperature dependence of ED differences due to variation in the C/O ratio.

For WASP-103b, the $P$-$T$ profiles as well as the chemical abundance profiles of the atmosphere are very different compared to WASP-101b and WASP-14b. The presence of temperature inversion in the $P$-$T$ profile being the major difference. At a solar C/O ratio similar atmospheric layers are probed in both the 3.6 and 4.5 $\micron$ channels. While emission due to CO from the inversion layer contributes to the 4.5 $\micron$ channel, VO from the inversion layer also contributes to 3.6 $\micron$ channel \citep[see the chemical abundance plot in the third row of Figure~\ref{fig:theory} and absorption cross-section Figure 2b in][]{Goyal2020}. However, the lower abundance of VO, combined with differences in the absorption cross-sections of VO and CO, lead to consistently lower EDs in the 3.6 $\micron$ band compared to the 4.5 $\micron$ band across a wide range of parameter space (as seen in the ED plot in the first row of Figure~\ref{fig:theory}). The only exception is the vertical branch at $\sim$2000 ppm in the 3.6 $\micron$ channel, where the change in metallicity at a recirculation factor of 0.25 alters the CO abundances (similar to WASP-101b) and therefore the 4.5 $\micron$ band ED. At a C/O ratio of 1.0, deeper pressures are probed in both of the channels due to the change in the $P$-$T$ profile, however, much deeper pressures are probed in the 3.6 $\micron$ channel (unlike WASP-101b and WASP-14b) since the abundance of VO decreases while that of CO approximately remains the same. The increase in the abundances of species, such as CH$_4$ and HCN, for higher C/O ratios is lower for WASP-103b compared to WASP-101b and WASP-14b, leading to smaller percentage variations in the 3.6 $\micron$ ED with the C/O ratio.  

In summary, the emission spectra in \textit{Spitzer} IRAC channel 2 (4.5 $\micron$) is dominated by CO across the parameter space for all the planets. This leads to minor variations in the ED as the C/O ratio changes. However, larger ED variations occur when metallicity changes, due to the sensitivity of the CO abundance to metallicity. In \textit{Spitzer} IRAC channel 1 (3.6 $\micron$) many other species (e.g. CH$_4$, HCN and VO) can shape the spectrum, depending on the C/O ratio, thus leading to a large spread in the 3.6 $\micron$ band ED with the C/O ratio for all the planets (see Figure~\ref{fig:theory}). The percentage variations in ED in both of the bands strongly depends on the temperature (via the recirculation factor and the $P$-$T$ profile), such that planets with lower equilibrium temperatures show increased percentage variations with C/O ratio and metallicity, while the percentage variations are smaller for planets with higher equilibrium temperatures (e.g. WASP-103b).

The observations from G20 and B20 are also shown in Figure~\ref{fig:theory} for each of the planets. For WASP-101b the observations are consistent with the f$_{\textup{c}}$ value of 1 (no heat redistribution) with the lower 1$\sigma$ range extending to  f$_{\textup{c}}$ = 0.75 (see Figures \ref{fig:obs} and \ref{fig:app22}). For the metallicity we do not obtain any constraints and a large range is possible all the way from solar to 200 times solar metallicity. However, the C/O ratio is constrained between a subsolar value of (0.35) to a slightly super-solar value of 0.75. For WASP-14b the observations are consistent with the f$_{\textup{c}}$ value of 0.75 (no heat redistribution) with the upper 1$\sigma$ range extending to  f$_{\textup{c}}$ = 1.0. The metallicity and C/O ratio are totally unconstrained for WASP-14b. For WASP-103b the observations are consistent with the f$_{\textup{c}}$ value of 1 (no heat redistribution) with the lower 1$\sigma$ range extending to  f$_{\textup{c}}$ = 0.75. The metallicity and C/O ratio are totally unconstrained even for WASP-103b. These constraints for the population of planets are discussed in Section \ref{subsec:results3}. We provide ED plots for all of the planets in our analysis, similar to top panel of the Figure~\ref{fig:theory}, in the figure set of Figure 10 in Appendix \ref{app:eclipse_plots} in the online version of the journal.

\begin{figure*}
\centering
\includegraphics[width=\textwidth]{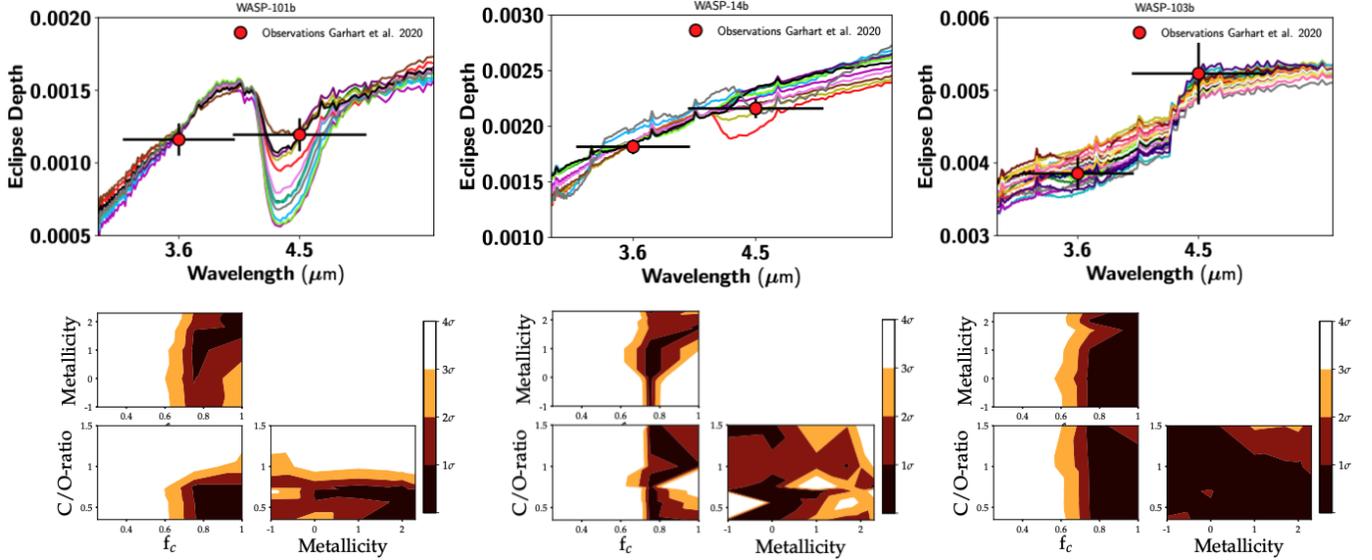}
\caption{Top: figures showing all the model emission spectra for WASP-101b, WASP-14b and WASP-103b (same planets as in Figure~\ref{fig:theory}) within 1$\sigma$ of the best fit model spectra when compared with observations (red points with error bars) from G20. Bottom: $\chi^2$ map of each grid parameter (f$_{\textup{c}}$, metallicity and C/O ratio) for WASP-101b, WASP-14b and WASP-103b observations from G20, fitted to all the model simulated spectra from the planet-specific grid for each planet. 1--4$\sigma$ contours with respect to the best-fit model parameters are shown.}
\label{fig:best_fit_chi_map}
\end{figure*}

\subsection{Comparison of Observations with Model Spectra}
\label{subsec:results3}

\begin{figure*}[ht!]
\centering
\includegraphics[width=\textwidth]{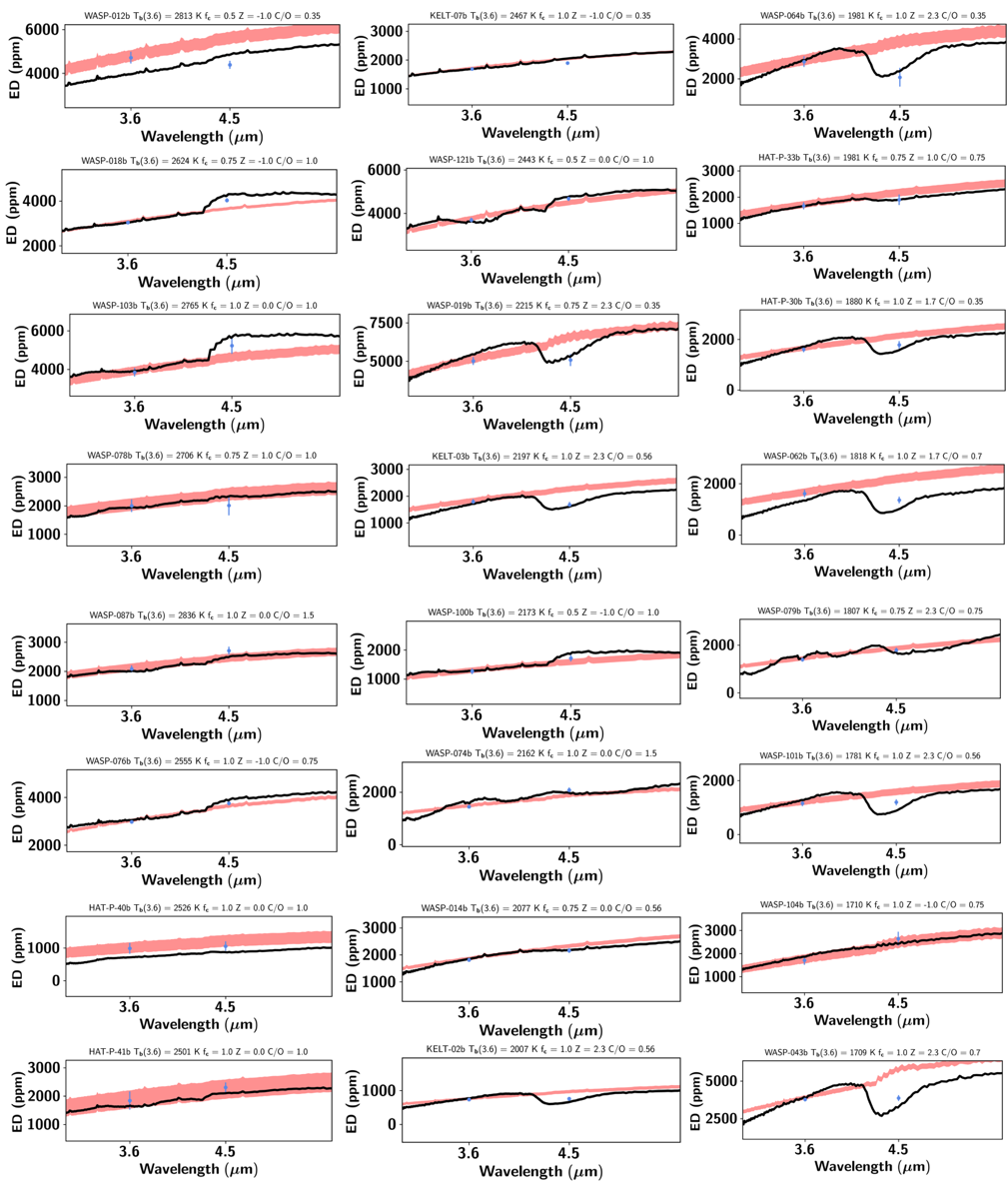}
    \label{fig:app21}
\end{figure*}

\begin{figure*}
\centering
\includegraphics[width=\textwidth]{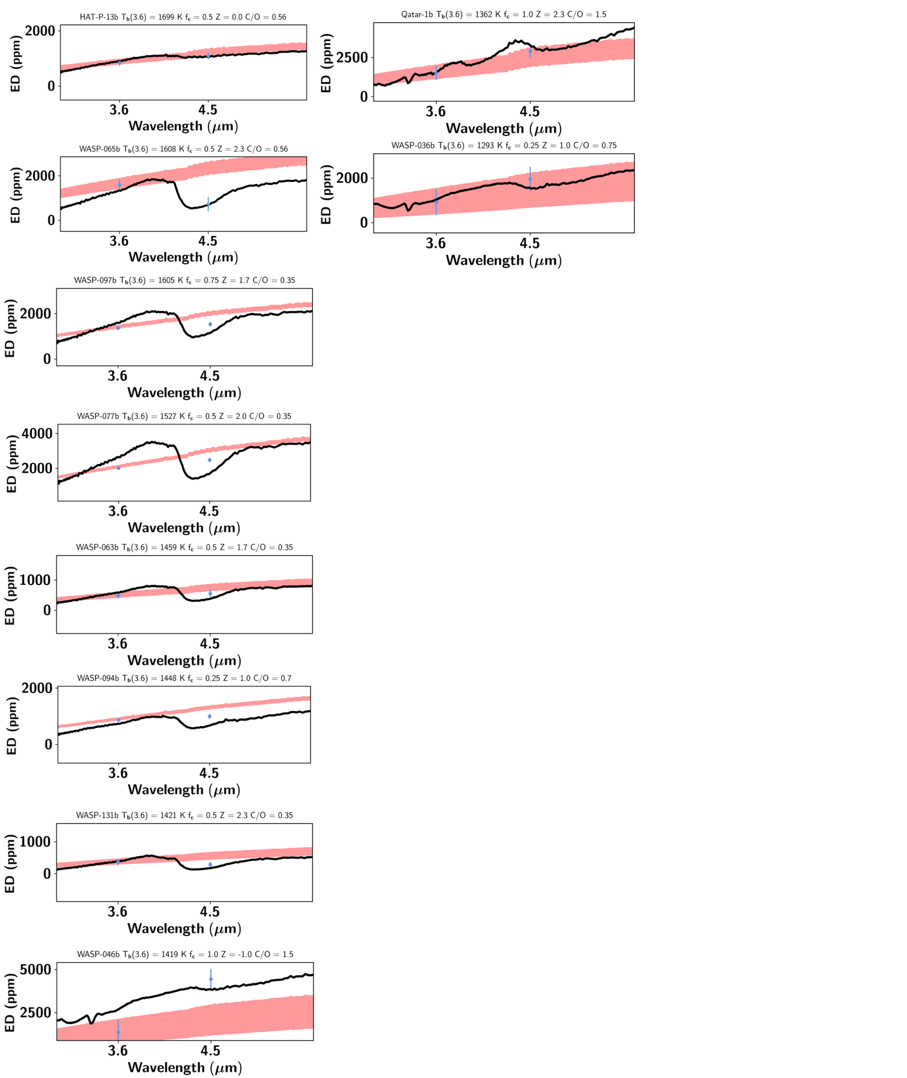}
 \caption{The best-fit emission spectra (black) for 34 planets from our planet-specific self-consistent grid are compared to the \textit{Spitzer} IRAC channel 1 and 2 EDs from G20. The red filled curve shows the blackbody emission spectrum for each planet using their observed channel 1 (3.6 $\micron$) brightness temperature calculated in this work as described in Section \ref{sec:bt}. The planets are arranged in decreasing order of their brightness temperature from top to bottom and left to right. We conclude that most hot Jupiters are inconsistent with blackbodies, in agreement with previous works \citep{Garhart2020,Baxter2020}. The best-fitting model parameters for each planet -- recirculation factor (f$_{\textup{c}}$), metallicity (Z) and C/O ratio -- are annotated at the top of each panel.}.
    \label{fig:app22}
\end{figure*}

\begin{figure*}
\centering
\includegraphics[width=\textwidth]{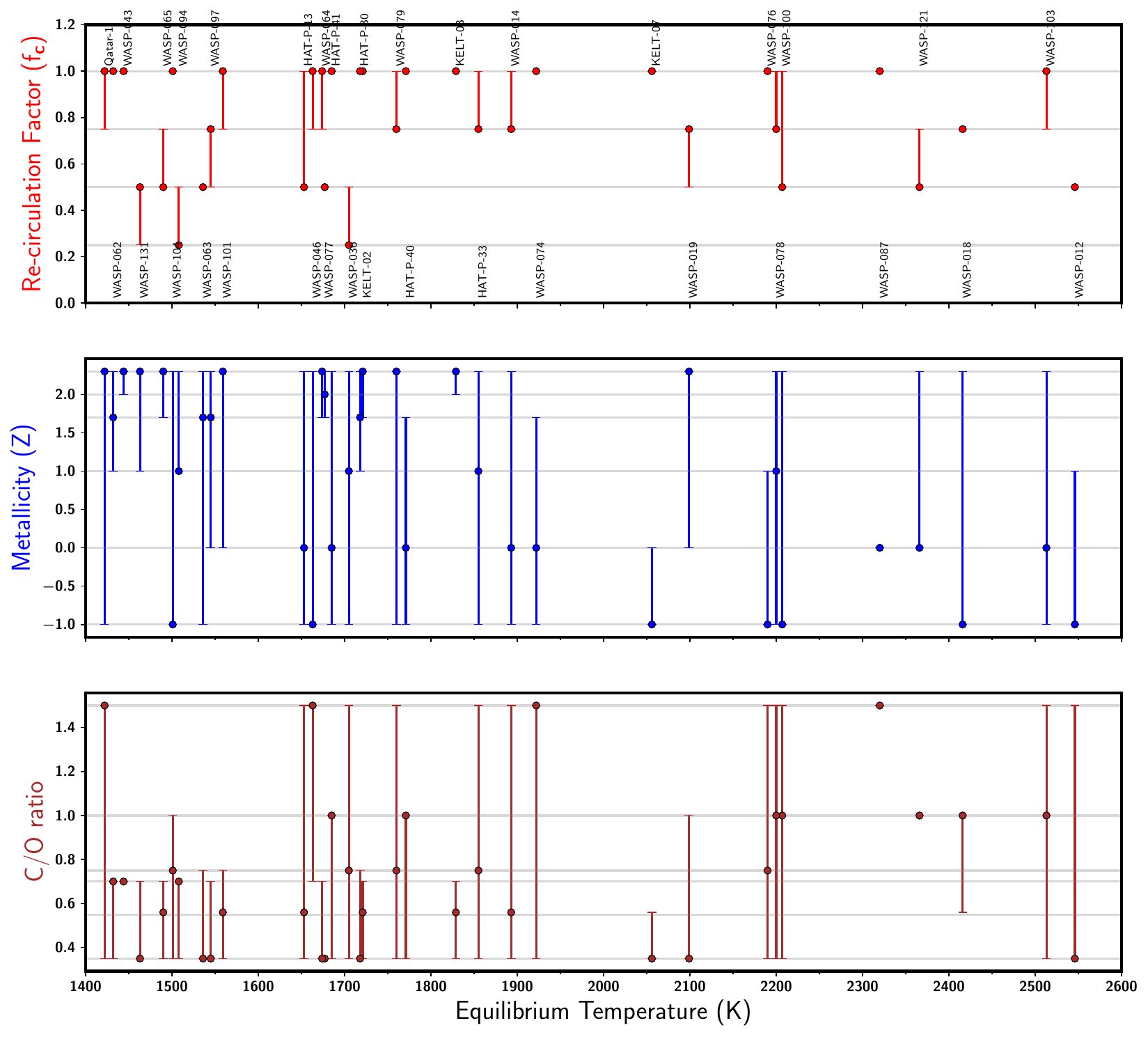}
\caption{Atmospheric parameter trends derived from \textit{Spitzer} EDs of hot Jupiters. Each planet is uniquely represented by its T$_{\textup{eq}}$ on the x-axis. The best-fitting (filled points) model parameters (minimum $\chi^2$), and their $1\,\sigma$ error bars (model parameters within the $1\,\sigma$ $\chi^2$ of the best-fit model), with increments of grid spacing for each parameter (gray horizontal lines), are shown for the recirculation factor (top panel), log$_{10}$ of metallicity with respect to solar metallicity (middle panel) and the C/O ratio (bottom panel) as a function of T$_{\textup{eq}}$. Planet names are annotated in the top panel which can be used to read their constraints of all the 3 model parameters. The associated best-fitting spectra are shown in Figure~\ref{fig:app22}. The large population of planets generally favor high f$_{\textup{c}}$ (f$_{\textup{c}}$ $\geq$ 0.75) values and C/O ratios $\leq$1. A broad range of metallicities are possible, without any population-level trends. A broader range of f$_{\textup{c}}$ is able to match the data for planets with T$_{\textup{eq}}$ below $\sim$1800\,K.}
\label{fig:obs}
\end{figure*}

We now compare our self-consistent planet-specific grid of models to all the planets with ED observations in G20. As an example, Figure~\ref{fig:best_fit_chi_map} (top panel) shows the \textit{Spitzer} observations of WASP-101b, WASP-14b and WASP-103b (the same set of planets as in Figure~\ref{fig:theory}) and all the models within $1\,\sigma$ of the minimum $\chi^2$ (i.e. best-fitting) model. We show in Figure~\ref{fig:best_fit_chi_map} (bottom panel) the corresponding $\chi^2$ map of each grid parameter for the observations from G20, fitted to all the model simulated spectra from the planet-specific grid for each planet with 1-4$\sigma$ contours \citep[see][for more details]{Goyal2020}. The $\chi^2$ map reveals that the \textit{Spitzer} observations of all three planets favor models with high recirculation factors (f$_{\textup{c}}$ $\geq$ 0.75 within the 1$\sigma$ contour). Moreover, WASP-101b exhibits a preference for low C/O ratios (i.e., less than 0.75 within 1$\sigma$ and less than 1.0 within 3$\sigma$ of the best-fit model). However, we are unable to place any tight constraints on the metallicity or the C/O ratio for these three planets, given the precision and wavelength coverage of these \textit{Spitzer} observations. This can also be noticed in Figure~\ref{fig:obs}, where many models with different metallicities and C/O ratios lie within $1\,\sigma$ of the best-fitting model. 

We conducted a similar analysis for the full hot Jupiter population covered in G20. A compendium of our best-fitting self-consistent model emission spectra to the \textit{Spitzer} IRAC channel 1 and 2 data for this population in shown in Figure~\ref{fig:app22}. For comparison, we also show the equivalent blackbody curve derived from the observed T$_{\textup{b}}$ of channel 1 alone (via Equation~\ref{eq:BT_model3}). We conclude that our self-consistent models generally achieve much better fits to the \textit{Spitzer} IRAC observations than blackbodies; that is, hot Jupiters do not emit like blackbodies, in agreement with previous works \citep{Triaud2014, Garhart2020, Baxter2020}. Moreover, the CO feature at 4.5\,$\micron$ can be clearly identified for many of the planets either in emission or absorption. Going from high to low temperature planets (high to low T$_{\textup{b}}$) a general trend is seen where CO is seen as an emission feature for high temperature planets due to the presence of temperature inversion in their $P$-$T$ profile, while CO/CO$_2$ is seen as an absorption feature in many of the low temperature planets without the temperature inversions. However, there are some exceptions depending on the best-fitting model parameters. For example, even though WASP-19b is a high temperature planet, with the potential to form a temperature inversion with an associated CO emission feature, our model fit to the data suggests that CO is seen as an absorption feature. Indeed, the best-fit model parameters for WASP-19b lead to a self-consistent $P$-$T$ profile without a temperature inversion. Due to such exceptions caused by the dependence on metallicity and the C/O ratio,  we do not see any clear transition with T$_{\textup{eq}}$ as in B20.

For model fits of all the planets shown in Figure~\ref{fig:app22} we also compute the 1$\sigma$ value of the model parameters with respect to the best-fit model as done for WASP-101b, WASP-14b and WASP-103b in Figure~\ref{fig:best_fit_chi_map}. Our constraints on the model parameters, i.e., atmospheric recirculation factor ( f$_{\textup{c}}$), metallicity, and C/O ratios for this population of planets are summarized in Figure~\ref{fig:obs}. Note that each planet is uniquely represented by its T$_{\textup{eq}}$ on the x-axis. The error bars indicate one sigma parameter value of the model with respect to the best-fit model. The observations of 62 \% of the planets from the sample are consistent with f$_{\textup{c}}$ $\geq$ 0.75, while  91 \% of the planets from the sample are consistent with f$_{\textup{c}}$ $\geq$ 0.5. Thus, a preference for inefficient recirculation of energy from the dayside to nightside of the planet (high f$_{\textup{c}}$ values) can be inferred for the population of planets from this figure (top panel). Above $\sim$1800\,K, the range of allowed f$_{\textup{c}}$ is much more narrow and skewed toward inefficient recirculation of energy (high f$_{\textup{c}}$ values). Below $\sim$1800\,K the range of allowed f$_{\textup{c}}$ values increases, with some of the lowest values for f$_{\textup{c}}$ (more efficient recirculation of energy from the dayside to the night side) allowed at low T$_{\textup{eq}}$s. This trend is consistent with theoretical expectations from the relative magnitudes of the radiative and advective timescales as one progresses from higher to lower T$_{\textup{eq}}$ planets \citep[e.g., see][]{Cowan2011, Perez-Becker2013, Schwartz2015}. It is important to note that the points in Figure~\ref{fig:obs} with no error bars aren't infinitely well constrained parameters, it is just that the neighboring values of those parameters in the grid are not within the 1$\sigma$ $\chi^2$ of the best-fit model. 

We were unable to obtain any population-level trends for the metallicity of the planetary atmospheres (middle panel in Figure~\ref{fig:obs}). A broad range of metallicities is possible for planets, from 0.1 times solar all the way up to 200 times the solar value. For the C/O ratio also a wide range from subsolar value (0.35) to super-solar value (1.5) are possible for the population of planets shown in Figure~\ref{fig:obs}. However, about 59 \% of the planets tend to favour C/O ratio $\leq$ 1.0, while 35 \% of the planets span a broad range of C/O ratio anywhere between 0.35 and 1.5.

We are not able to identify any definitive population-level trends for metallicity and the C/O ratio, and also trends in these parameters with respect to planetary T$_{\textup{eq}}$ solely from \textit{Spitzer} observations. One of the main reasons for this being the larger error bars in current observations, in comparison to the change in the 3.6 and 4.5 $\micron$ EDs due to metallicity and the C/O ratio. The other reason being the lack of spectral resolution and range. Our results for the C/O ratio are inline with the findings of B20, where they conclude that hot Jupiters with high C/O ratios (C/O $\geq$ 0.85) are rare. However, it is difficult to make a definitive statement with the precision, spectral resolution and range of the current dataset. We note the our sample size for this analysis is from G20 (34 planets with positive ED values) which is smaller than the sample size of B20. 

\begin{figure*}
\centering
\includegraphics[width=1.0\textwidth]{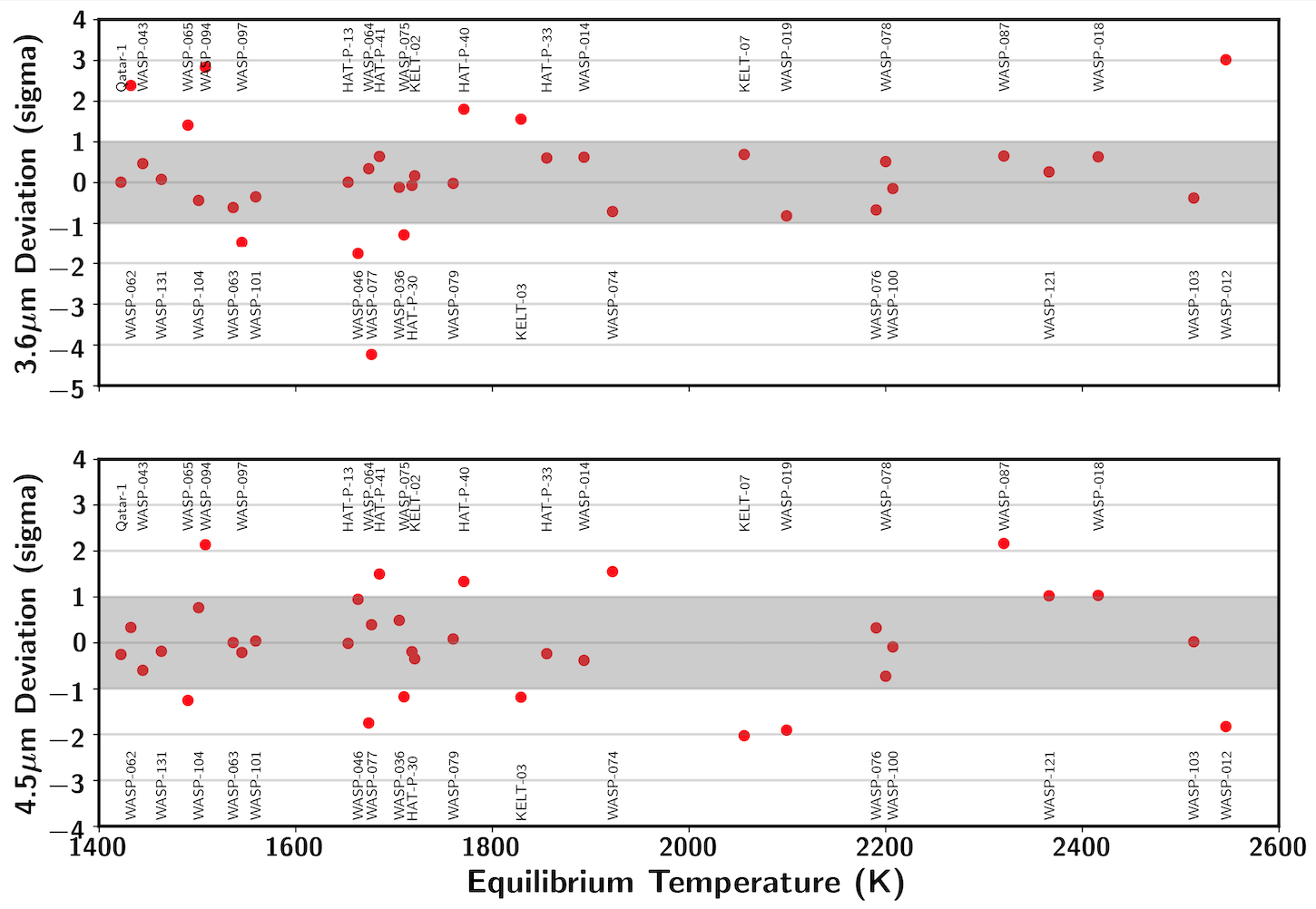}
\caption{Figure showing the deviation (in sigma units) of the model EDs in comparison to observed EDs in \textit{Spitzer} 3.6 $\micron$ IRAC 1 band (top panel) and 4.5 $\micron$ IRAC 2 band (bottom panel) for all of the planets considered. Each planet is uniquely represented by its T$_{\textup{eq}}$ on the x-axis, while also alternately annotated on the upper and lower edge of each panel from low to high T$_{\textup{eq}}$. The gray color shaded area shows the $\pm$1 $\sigma$ deviation.}
\label{fig:obs_model}
\end{figure*}

\begin{figure*}[ht!]
\subfloat[]{\includegraphics[width=\columnwidth]{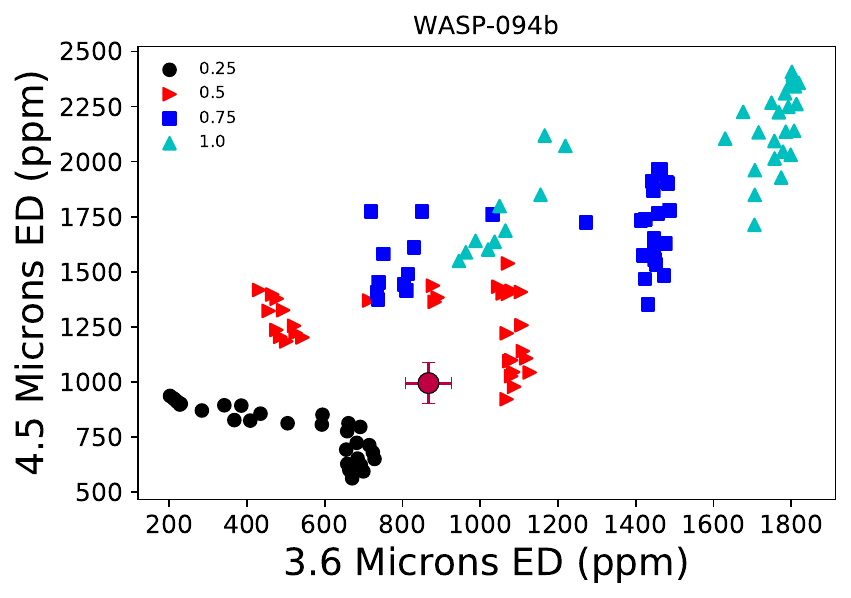}\label{fig:ana_WASP-94}}
\subfloat[]{\includegraphics[width=\columnwidth]{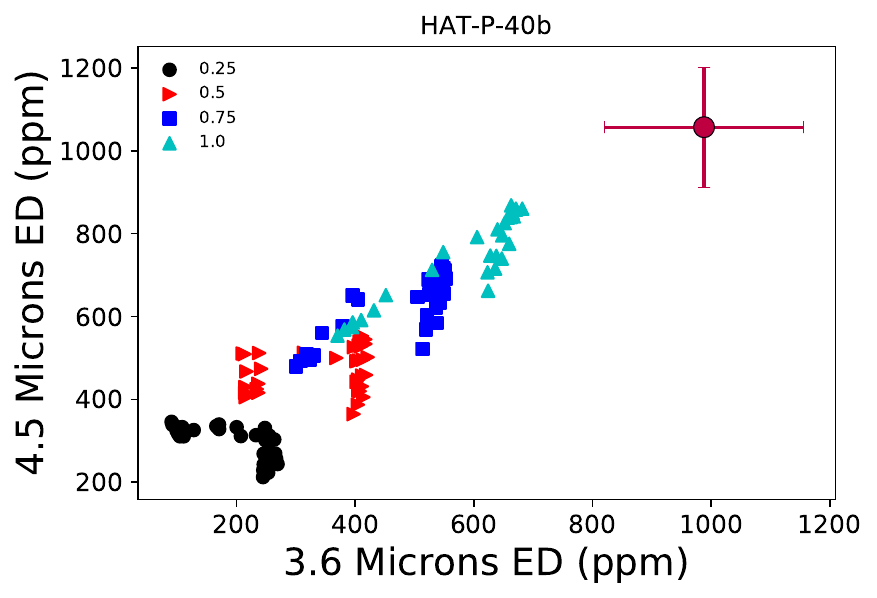}\label{fig:ana_HAT-P_40}}\\
\subfloat[]{\includegraphics[width=\columnwidth]{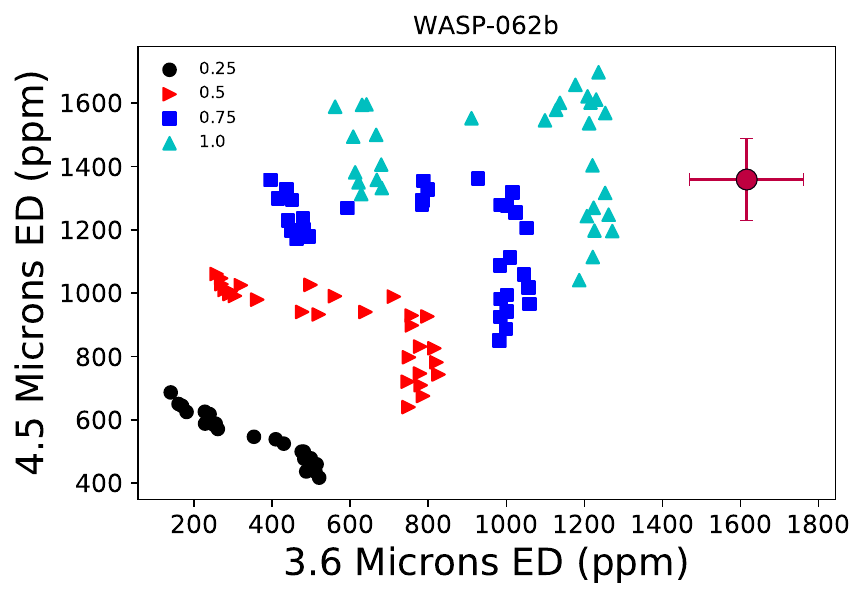}\label{fig:ana_WASP-62}}
\subfloat[]{\includegraphics[width=\columnwidth]{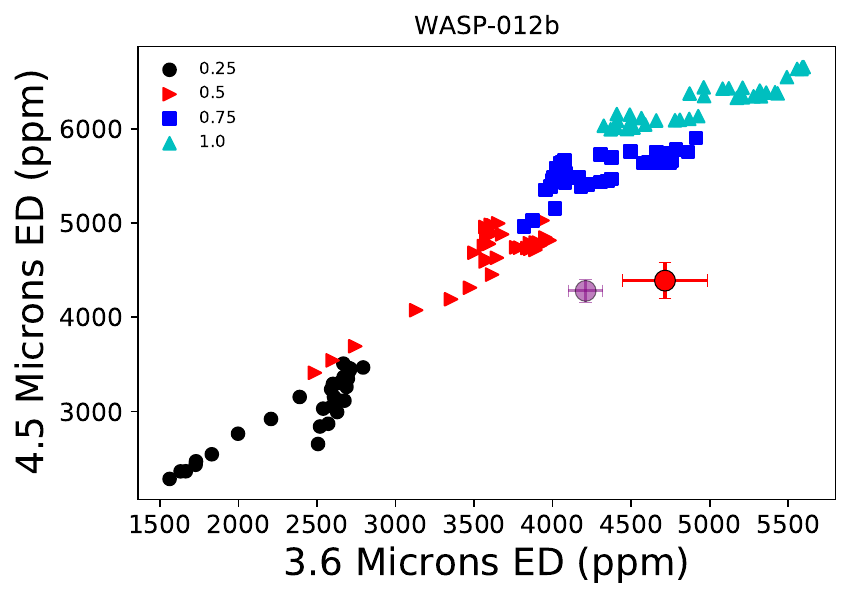}\label{fig:ana_WASP-12}}
\caption{\textbf{(a)} The \textit{Spitzer} channel 1 and 2 model EDs for WASP-94b across the entire planet-specific grid, compared to the observed ED from G20 and B20 (they are overlapping). A model with 0.25 $\ge$ f$_{\textup{c}}$ $\le$ 0.5 should be able to explain the observations. \textbf{(b)} Similar to \ref{fig:ana_WASP-94}, but for HAT-P-40b. The observed ED is substantially higher that the hottest (largest ED) model in our grid for this planet in both the channels. \textbf{(c)} Similar to \ref{fig:ana_WASP-94}, but for WASP-62b. Here, the observed channel 1 ED is much higher than the model with largest channel 1 ED. \textbf{(d)} Similar to \ref{fig:ana_WASP-94}, but for WASP-12b. Here, observed ED for both the channels lie outside the theoretical trend structure of all the model simulations in the grid.}
\end{figure*}

\subsection{Discrepancies between models and observations}
\label{subsec:results4}

For most of the planets in G20, we find at least a subset of our models provide a good fit (any of the model simulations from the grid lie within the error bars of the observations) to the observations. In Figure~\ref{fig:obs_model}, we show the deviations of model EDs from the observed EDs in each of the \textit{Spitzer} IRAC channels for all the planets from G20 considered in this work. It can be noticed that for most of the planets 3.6 and 4.5 $\micron$ model EDs lie within $\pm$1$\sigma$ of the observed EDs. However, there are some planets for which the model spectrum is not a good fit (see Figure~\ref{fig:app22}) and the model ED lie outside the 1$\sigma$ error bar of the observations, in either of the channels. Many of these fits are not good because the model grid has its own resolution (spacing) between each parameter (e.g. 0.25 in f$_{\textup{c}}$). Therefore, by increasing the model grid resolution of each parameter, one can provide a better fit. For example, the observations of WASP-94b could be possibly explained by a model simulation with f$_{\textup{c}}$ between 0.25 and 0.5, as seen in Figure~\ref{fig:ana_WASP-94}. However, the \textit{Spitzer} IRAC observations for some of the planets cannot be fit by our models, even at the extreme edges of our grid parameter space. This effect is particularly strong for HAT-P-40b, WASP-62b and WASP-12b as shown in Figure~\ref{fig:ana_HAT-P_40}, \ref{fig:ana_WASP-62} and \ref{fig:ana_WASP-12}, respectively. These anomalies also exist, but to a lesser extent, for a few other planets in our analysis, particularly, WASP-65b and WASP-87b. We provide ED plots, similar to Figure~\ref{fig:ana_HAT-P_40}, for all the planets included in our analysis in the online version of the journal in Appendix \ref{app:eclipse_plots}.

As shown in Figure~\ref{fig:ana_HAT-P_40} for HAT-P-40b (and WASP-87b), the observed EDs in both the \textit{Spitzer} channels are larger than the prediction of the simulated model spectra with the largest ED (hottest models with f$_{\textup{c}}$ = 1) across the entire planet-specific grid. This indicates that the atmosphere of HAT-P-40b is substantially hotter than the models for which the entire host star energy is dumped on the dayside of the planet. For WASP-62b shown in Figure~\ref{fig:ana_WASP-62} (and WASP-65b), the observed ED only in channel 1 (3.6 $\micron$) is larger than the largest channel 1 model ED (hottest model in channel 1) across the entire planet-specific grid. For WASP-12b observed channel 1 and 2 ED is not higher than the largest model ED, but lies outside the theoretical trends of all the model simulations in the grid as shown in Figure~\ref{fig:ana_WASP-12}. The discrepancy in the observations between G20 and B20 can also be noted for WASP-12b in Figure~\ref{fig:ana_WASP-12} and Appendix \ref{app1}. For the planets where we see anomalies between observations and models, most commonly the observed channel 1 ED (3.6 $\micron$) is larger (hotter) than the model ED. This happens for fewer planets in channel 2 (4.5 $\micron$). This motivates us to investigate the mechanisms that could cause such anomalies, especially in channel 1.  

We first assessed the potential role of an unknown opacity source in causing these anomalies by adding/removing a grey absorbing opacity in either or both of \textit{Spitzer} channels 1 and 2. We find that for planets with low T$_{\textup{eq}}$ (and hence without inversions), such as WASP-62b, removing (decreasing) opacity in a given band tends to increase the model ED. This is due to deeper, higher temperature, layers of the atmosphere being probed, thus taking the model ED closer to observed ED shown in Figure~\ref{fig:ana_WASP-62}. However, for planets with a temperature inversion in their $P$-$T$ profile such as WASP-12b, adding gray opacity in a given band tends to increase the ED in that band, thus bringing our model EDs closer to the observed value. Therefore, the presence/absence of some opacity source in our models that may or may not be present in the atmosphere of the planet could be one of the potential reasons for the anomalies we see for some planets. The opacity in a given band is also a function of the abundance of the species that has strong absorption cross-section in that band. Therefore, another reason for the anomaly could be that some of the chemical species that are important in either of the bands, such as CH$_4$ in channel 1 and CO in channel 2, may have disequilibrium abundances leading to higher/lower EDs. For example, we tested increasing the CH$_4$ abundance to $\sim 10^{-4}$, that is $\sim 10^5$ times its solar value for WASP-12b at 1 bar, in which case the model simulation was able to explain the anomalously high channel 1 (3.6 $\micron$) ED seen in the observations. The detailed analysis of these anomalies is beyond the scope of this study and will be addressed in our future work considering nonequilibrium chemistry. As we noted earlier, the model simulations in this grid are cloud-free, however, the presence of clouds leads to emission from lower pressures within an atmosphere and hence lower temperature. This will only lead to a decrease in the ED. Therefore, even the presence of clouds cannot explain this enhanced observed ED (hotter temperatures) for certain planets. We also note that we do not include opacity due to many of the refractory species such as Fe II, Ti I, Ni I, Mn I, Ca I, Ca II etc. in our model. These species can substantially increase transit depths in the optical part of the spectrum (due to their strong UV-optical opacities), especially for ultra-hot Jupiters \citep{Lothringer2020, Lothringer2021}. Therefore, these species have the potential to effect the $P$-$T$ profile, and thereby the emission spectrum in the \textit{Spitzer} IRAC channels. These atomic and ionic metallic species could therefore be one source of the anomaly we identify in this work for ultra-hot Jupiters.

\section{Conclusions}\label{sec:conclusions}

In this study, we considered theoretical explanations for thermal emission observations of 34 hot Jupiters via \textit{Spitzer} IRAC channel 1 (3.6\,$\micron$) and channel 2 (4.5\,$\micron$). We first benchmarked our emission spectra model with a published model in the literature. We highlighted subtle differences in computing brightness temperatures (T$_{\textup{b}}$) using observed EDs compared to computing EDs directly from a model spectrum. We conclude that T$_{\textup{b}}$ and their uncertainties are highly influenced by modeler choices, which can lead to large variations where assumptions differ. We therefore advocate using EDs to directly compare observations with models. Our main results are as follows:

\begin{enumerate}
    \item We presented theoretical trends in EDs in \textit{Spitzer} channels 1 and 2 for a population of hot Jupiters, spanning a range of recirculation factors, metallicities and C/O ratios. We find that the emission spectrum in \textit{Spitzer} IRAC channel 2 (4.5 $\micron$) is dominated by CO across the parameter space, leading to minor variations in the ED due to the C/O ratio. However, changes in metallicity lead to large variations in the ED in this channel, driven by changes in the abundance of CO with metallicity. In contrast, many other species (e.g. CH$_4$, HCN and VO) can shape the spectrum in \textit{Spitzer} IRAC channel 1 (3.6 $\micron$ band), depending on the C/O ratio and $P$-$T$ profile, thus leading to a large spread in ED in this band with the C/O ratio. The percentage variations in ED in both of the bands has a strong dependence on temperature (via recirculation factor and consistent $P$-$T$ profiles), such that planets with lower equilibrium temperatures show a higher percentage of variations with the C/O ratio and metallicity, while the percentage of variations is smaller for planets with higher equilibrium temperatures.

    \item Our best-fitting model spectra for each of the planets demonstrate that \textit{Spitzer} observations of hot Jupiters are poorly described by blackbodies. This is consistent with the findings of the previous studies \citep{Triaud2014,Garhart2020,Baxter2020}. The spectral fits for a population of planets show that the planets with high equilibrium temperatures tend to show CO emission features (in \textit{Spitzer} channel 2) indicating temperature inversions, while planets with low equilibrium temperatures show CO/CO$_2$ absorption features indicating a $P$-$T$ profile without a temperature inversion. However, 
    there are many exceptions governed by the metallicity and C/O ratio of the best-fit model atmosphere. Therefore, 
    we do not see any clear transition with equilibrium temperature as concluded in \citet{Baxter2020}.
    
    \item By comparing our self-consistent models with the \textit{Spitzer} observations from \citet{Garhart2020}, we find a trend in the recirculation factor (f$_{\textup{c}}$) across the hot-Jupiter population. The majority of hot Jupiters favor higher values of f$_{\textup{c}}$, indicating inefficient recirculation of energy from their dayside to nightside (hotter dayside than expected with efficient recirculation) in agreement with previous findings for different populations of planets \citep{Cowan2011,Schwartz2015}. Specifically, for equilibrium temperatures above $\sim$1800\,K the range of allowed f$_{\textup{c}}$ is much more narrow and skewed toward inefficient recirculation (f$_{\textup{c}} > 0.5$). We also see that 59 \% of our sample size of hot Jupiters are consistent with a C/O ratio of less than 1 and 35 \% are consistent with the whole range of C/O ratio (0.35 $\leq$ C/O $\leq$ 1.5). We do not see any definitive population-level trend for metallicity and the C/O ratio. This is because the variation in EDs caused by metallicity and C/O ratio variations is smaller than the \textit{Spitzer} data precision, making them indistinguishable. The lack of spectral coverage and resolution in \textit{Spitzer} also make it difficult to place strong constraints on metallicity and the C/O ratio. We also do not see any trend in metallicity or C/O ratio with equilibrium temperature, most likely due to poor constraints. We also find that, for most of the planets in our sample, 3.6 and 4.5 $\micron$ model EDs lie within $\pm$1 $\sigma$ of the observed EDs.
    \item Some hot Jupiters, such as HAT-P-40b and WASP-62b, display substantially larger EDs than the maximum theoretical value predicted by our planet-specific grid. That is, some of the planets are significantly hotter than expected in one or both of the \textit{Spitzer} channels. Such anomalies could arise from the disequilibrium abundance of the species whose opacity dominate in that specific channel or due to opacity of any atomic/ionic species not included in our current model. This will be addressed in more detail in our future work.
\end{enumerate}



In the near future, the James Webb Space Telescope (JWST) will provide thermal emission spectra for many of these hot Jupiters with much higher spectral resolution and precision. The planet-specific grid of self-consistent atmospheres presented in this work may then be used to precisely constrain atmospheric metallicities and C/O ratios. We eagerly await the improved data quality from JWST emission spectra, which may illuminate the underlying cause of the anomalously high EDs of some of the hot Jupiters and give more stronger constraints on their atmospheric recirculation, metallicity and C/O ratio.

\acknowledgments
We thank Professor Jonathan Fortney for providing us with the relevant model spectrum from G20 for comparison with the \texttt{ATMO} models utilized in this study. JMG and NKL acknowledge funding from NASA grant 80NSSC20K0586 from JWST GTO program. NJM would like to acknowledge funding from a Science and Technology Facilities Council Consolidated grant (ST/R000395/1), the Leverhulme Trust through a research project grant RPG-2020-82 and the UKRI Future Leaders Scheme: MR/T040866/1.

%







\appendix

\section{Table of Observed and Derived Quantities for Each Planet.}
\label{app1}
Table \ref{target_table} summarizes the equilibrium temperature, \textit{Spitzer} IRAC channel 1 and 2 EDs and T$_{\textup{b}}$ from G20 and B20, and T$_{\textup{b}}$ derived in this work (JG21), for all the hot-Jupiter exoplanets with positive values of EDs from G20. 
\movetabledown=25mm
\begin{rotatetable}
\begin{deluxetable}{cccccccccccc}
\tablecaption{Observed Eclipse Depth (ED) and Derived Brightness Temperature (T$_{\textup{b}}$) for Each Planet from Different Works}
\centerwidetable
\tabletypesize{\footnotesize} 
\tablecolumns{12}
\tablewidth{15pt}
 \tablehead{
 \colhead{Planet} &  \colhead{Equilibrium} &	 \colhead{G20 3.6 ED} &  \colhead{G20 4.5 ED} &  \colhead{B20 3.6 ED} &  \colhead{B20 4.5 ED}	&  \colhead{G20 3.6 T$_{\textup{b}}$} &  \colhead{G20 4.5 T$_{\textup{b}}$} &  \colhead{B20 3.6 T$_{\textup{b}}$} &  \colhead{B20 4.5 T$_{\textup{b}}$} &  \colhead{JG21 3.6 T$_{\textup{b}}$} &  \colhead{JG21 4.5 T$_{\textup{b}}$} \\
 \colhead{Name} & \colhead{Temperature (K)} & \colhead{(ppm)} &  \colhead{(ppm)} &  \colhead{(ppm)} &  \colhead{(ppm)}	&  \colhead{(K)} &  \colhead{(K)} &  \colhead{(K)} &  \colhead{(K)} &  \colhead{(K)} &  \colhead{(K)}
}
\startdata
HAT-P-13	&	1653$\pm$50	&	851$\pm$107	&	1090$\pm$124	&	851$\pm$107	&	1090$\pm$124	&	1810$\pm$229	&	1754$\pm$200	&	1775$\pm$87	&	1728$\pm$89	&	1788$\pm$89	&	1735$\pm$89	\\
HAT-P-30	&	1718$\pm$34	&	1603$\pm$108	&	1783$\pm$149	&	1584$\pm$107	&	1825$\pm$147	&	2087$\pm$140	&	1938$\pm$160	&	1868$\pm$51	&	1763$\pm$65	&	1894$\pm$52	&	1753$\pm$67	\\
HAT-P-33	&	1855$\pm$148	&	1663$\pm$132	&	1896$\pm$206	&	1603$\pm$127	&	1835$\pm$199	&	2112$\pm$162	&	1990$\pm$209	&	2000$\pm$67	&	1901$\pm$98	&	2034$\pm$70	&	1926$\pm$101	\\
HAT-P-40	&	1771$\pm$38	&	988$\pm$168	&	1057$\pm$145	&	988$\pm$168	&	1057$\pm$145	&	2074$\pm$354	&	1887$\pm$259	&	2005$\pm$146	&	1840$\pm$119	&	2022$\pm$148	&	1848$\pm$119	\\
HAT-P-41	&	1685$\pm$58	&	1842$\pm$321	&	2303$\pm$179	&	1829$\pm$319	&	2278$\pm$177	&	1694$\pm$294	&	1622$\pm$125	&	2173$\pm$172	&	2179$\pm$88	&	2196$\pm$174	&	2199$\pm$90	\\
KELT-02	&	1721$\pm$36	&	739$\pm$43	&	761$\pm$53	&	650$\pm$38	&	678$\pm$47	&	1994$\pm$104	&	1782$\pm$111	&	1862$\pm$44	&	1679$\pm$52	&	1979$\pm$48	&	1777$\pm$57	\\
KELT-03	&	1829$\pm$42	&	1788$\pm$98	&	1677$\pm$105	&	1766$\pm$97	&	1656$\pm$104	&	2445$\pm$133	&	2132$\pm$133	&	2300$\pm$59	&	2006$\pm$62	&	2320$\pm$60	&	2017$\pm$63	\\
KELT-07	&	2056$\pm$31	&	1688$\pm$46	&	1896$\pm$57	&	N/A	&	N/A	&	2512$\pm$69	&	2415$\pm$73	&	N/A	&	N/A	&	2431$\pm$32	&	2340$\pm$38	\\
Qatar-1	&	1422$\pm$36	&	1511$\pm$455	&	2907$\pm$415	&	1490$\pm$510	&	2730$\pm$490	&	1410$\pm$425	&	1532$\pm$219	&	1374$\pm$153	&	1470$\pm$108	&	1391$\pm$136	&	1512$\pm$89	\\
WASP-012	&	2546$\pm$82	&	4715$\pm$270	&	4389$\pm$188	&	4210$\pm$110	&	4280$\pm$120	&	3329$\pm$172	&	2934$\pm$114	&	2872$\pm$40	&	2649$\pm$43	&	3077$\pm$217	&	2698$\pm$167	\\
WASP-014	&	1893$\pm$60	&	1816$\pm$67	&	2161$\pm$88	&	1870$\pm$70	&	2240$\pm$180	&	2302$\pm$85	&	2256$\pm$92	&	2248$\pm$39	&	2221$\pm$93	&	2229$\pm$37	&	2180$\pm$46	\\
WASP-018	&	2416$\pm$58	&	3037$\pm$62	&	4033$\pm$97	&	3000$\pm$200	&	3900$\pm$200	&	3057$\pm$63	&	3323$\pm$80	&	2990$\pm$109	&	3231$\pm$104	&	2903$\pm$45	&	4756$\pm$249	\\
WASP-019	&	2099$\pm$39	&	5016$\pm$259	&	5081$\pm$392	&	4830$\pm$250	&	5720$\pm$300	&	2451$\pm$127	&	2191$\pm$169	&	2326$\pm$57	&	2270$\pm$63	&	2388$\pm$59	&	2140$\pm$84	\\
WASP-036	&	1705$\pm$44	&	914$\pm$579	&	1953$\pm$545	&	913$\pm$578	&	1948$\pm$544	&	1336$\pm$844	&	1506$\pm$420	&	1300$\pm$267	&	1475$\pm$168	&	1306$\pm$270	&	1472$\pm$168	\\
WASP-043	&	1444$\pm$40	&	3773$\pm$138	&	3866$\pm$195	&	3460$\pm$130	&	3820$\pm$150	&	1781$\pm$65	&	1537$\pm$78	&	1664$\pm$24	&	1497$\pm$24	&	1751$\pm$24	&	1521$\pm$32	\\
WASP-046	&	1663$\pm$54	&	1360$\pm$701	&	4446$\pm$589	&	1360$\pm$701	&	4446$\pm$589	&	1435$\pm$740	&	2014$\pm$267	&	1394$\pm$241	&	1968$\pm$129	&	1449$\pm$257	&	2074$\pm$139	\\
WASP-062	&	1432$\pm$33	&	1616$\pm$146	&	1359$\pm$130	&	1616$\pm$146	&	1359$\pm$130	&	1955$\pm$177	&	1593$\pm$153	&	1906$\pm$71	&	1568$\pm$63	&	1924$\pm$72	&	1575$\pm$64	\\
WASP-063	&	1536$\pm$37	&	486$\pm$96	&	560$\pm$130	&	552$\pm$95	&	533$\pm$128	&	1547$\pm$308	&	1395$\pm$324	&	1573$\pm$97	&	1347$\pm$123	&	1534$\pm$106	&	1395$\pm$127	\\
WASP-064	&	1674$\pm$169	&	2859$\pm$270	&	2071$\pm$471	&	2859$\pm$270	&	2071$\pm$471	&	2135$\pm$202	&	1607$\pm$366	&	2102$\pm$87	&	1610$\pm$159	&	2133$\pm$90	&	1628$\pm$162	\\
WASP-065	&	1490$\pm$45	&	1587$\pm$245	&	724$\pm$318	&	1587$\pm$245	&	724$\pm$318	&	1833$\pm$284	&	1179$\pm$518	&	1781$\pm$108	&	1160$\pm$177	&	1791$\pm$109	&	1162$\pm$177	\\
WASP-074	&	1922$\pm$46	&	1446$\pm$66	&	2075$\pm$100	&	1446$\pm$66	&	2075$\pm$100	&	2049$\pm$94	&	2161$\pm$105	&	1997$\pm$39	&	2106$\pm$51	&	1989$\pm$39	&	2089$\pm$51	\\
WASP-076	&	2190$\pm$43	&	2979$\pm$72	&	3762$\pm$92	&	2645$\pm$63	&	3345$\pm$82	&	2669$\pm$57	&	2747$\pm$60	&	2411$\pm$28	&	2471$\pm$33	&	2576$\pm$32	&	2799$\pm$110	\\
WASP-077	&	1677$\pm$28	&	2016$\pm$103	&	2487$\pm$134	&	1845$\pm$94	&	2362$\pm$127	&	1786$\pm$84	&	1696$\pm$87	&	1685$\pm$32	&	1628$\pm$37	&	1763$\pm$36	&	1682$\pm$41	\\
WASP-078	&	2200$\pm$41	&	2001$\pm$218	&	2013$\pm$351	&	2001$\pm$218	&	2013$\pm$351	&	3034$\pm$331	&	2763$\pm$483	&	2787$\pm$160	&	2565$\pm$255	&	2813$\pm$205	&	2594$\pm$534	\\
WASP-079	&	1760$\pm$51	&	1394$\pm$88	&	1783$\pm$106	&	1394$\pm$88	&	1783$\pm$106	&	1959$\pm$125	&	1948$\pm$117	&	1893$\pm$49	&	1882$\pm$54	&	1899$\pm$50	&	1884$\pm$53	\\
WASP-087	&	2320$\pm$62	&	2080$\pm$127	&	2708$\pm$137	&	2077$\pm$127	&	2705$\pm$137	&	2802$\pm$172	&	2988$\pm$152	&	2687$\pm$85	&	2863$\pm$87	&	2714$\pm$85	&	3589$\pm$334	\\
WASP-094	&	1508$\pm$75	&	867$\pm$59	&	995$\pm$93	&	867$\pm$59	&	995$\pm$93	&	1385$\pm$95	&	1249$\pm$118	&	1527$\pm$36	&	1398$\pm$50	&	1525$\pm$36	&	1392$\pm$50	\\
WASP-097	&	1545$\pm$40	&	1359$\pm$84	&	1534$\pm$101	&	1359$\pm$84	&	1534$\pm$101	&	1772$\pm$111	&	1615$\pm$107	&	1727$\pm$40	&	1590$\pm$44	&	1759$\pm$42	&	1613$\pm$46	\\
WASP-100	&	2207$\pm$170	&	1267$\pm$98	&	1720$\pm$119	&	1267$\pm$98	&	1720$\pm$119	&	2306$\pm$180	&	2429$\pm$168	&	2216$\pm$79	&	2337$\pm$88	&	2228$\pm$78	&	2344$\pm$89	\\
WASP-101	&	1559$\pm$38	&	1161$\pm$111	&	1194$\pm$113	&	1161$\pm$111	&	1194$\pm$113	&	1723$\pm$166	&	1524$\pm$145	&	1680$\pm$61	&	1492$\pm$58	&	1692$\pm$61	&	1498$\pm$58	\\
WASP-103	&	2513$\pm$49	&	3854$\pm$226	&	5230$\pm$424	&	4458$\pm$383	&	5686$\pm$138	&	2993$\pm$150	&	3268$\pm$231	&	3114$\pm$149	&	3337$\pm$52	&	2900$\pm$148	&	4905$\pm$895	\\
WASP-104	&	1501$\pm$189	&	1709$\pm$195	&	2643$\pm$303	&	1709$\pm$195	&	2643$\pm$303	&	1716$\pm$197	&	1783$\pm$205	&	1734$\pm$76	&	1828$\pm$98	&	1720$\pm$75	&	1802$\pm$96	\\
WASP-121	&	2366$\pm$57	&	3685$\pm$114	&	4684$\pm$121	&	3150$\pm$103	&	4510$\pm$107	&	2490$\pm$77	&	2562$\pm$66	&	2358$\pm$36	&	2591$\pm$35	&	2428$\pm$36	&	2497$\pm$36	\\
WASP-131	&	1463$\pm$32	&	364$\pm$96	&	289$\pm$80	&	364$\pm$97	&	282$\pm$78	&	1397$\pm$369	&	1114$\pm$309	&	1361$\pm$115	&	1077$\pm$96	&	1369$\pm$116	&	1090$\pm$97	\\
\enddata
\label{target_table}
\end{deluxetable}
\end{rotatetable}


\section{Eclipse Depth Variation with Metallicity}
\label{app:metallicity_plot}
This section shows variation of \textit{Spitzer} IRAC channel 1 and 2 model EDs with metallicity and the recirculation factor (f$_{\textup{c}}$) for WASP-101b on the left side of Figure~\ref{appfig:metal}. Additionally, the chemical abundance profiles of important species at solar and 100 times solar metallicity are also shown on the right side of Figure~\ref{appfig:metal}.

\begin{figure*}
\centering
\includegraphics[width=1.0\textwidth]{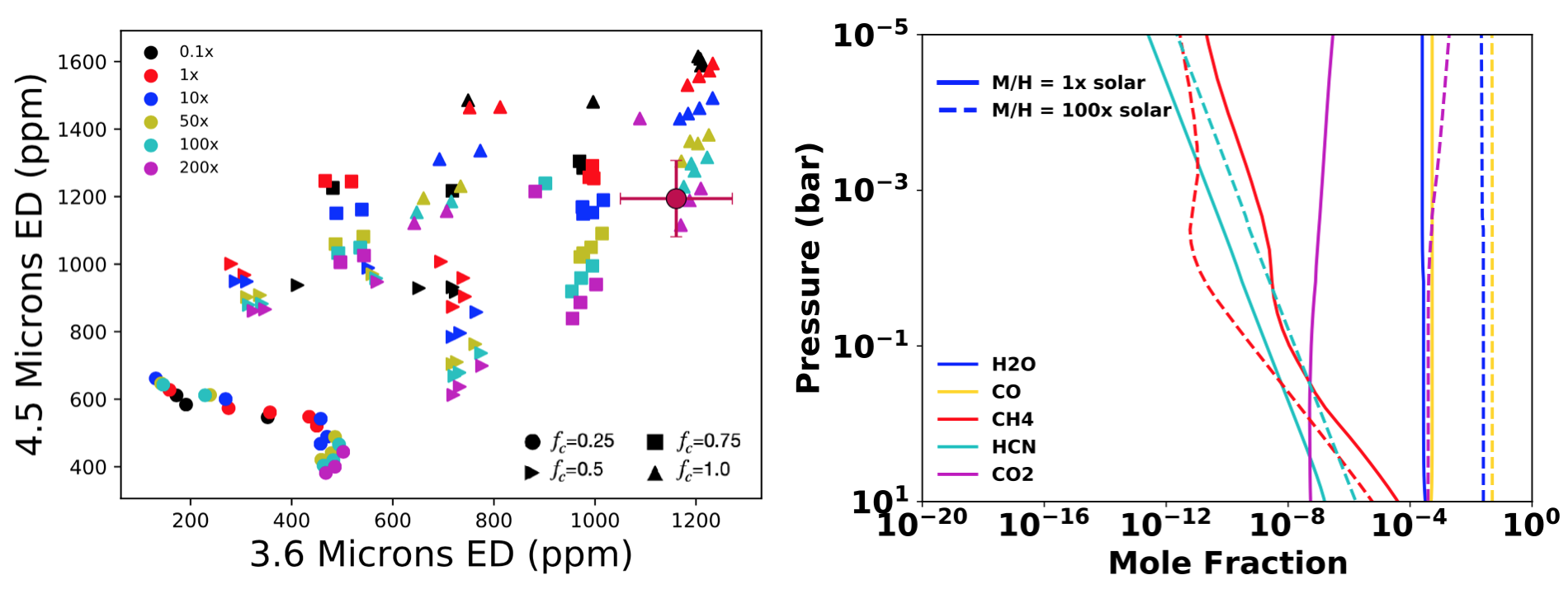}
\caption{Left: model EDs in the 3.6 and 4.5 $\micron$ \textit{Spitzer} bands for WASP-101b. The markers correspond to different f$_{\textup{c}}$ values, while the colors correspond to different metallicities (see the legends). The observed EDs from G20 and B20 are overlaid for comparison, which for WASP-101b from G20 and B20 overlap. Right: Chemical abundances at different atmospheric (pressure) layers for certain important chemical species with strong opacities in either of 3.6 and 4.5 $\micron$ \textit{Spitzer} bands, at solar metallicity (solid line) and 100 times solar metallicity (dashed line).}
\label{appfig:metal}
\end{figure*}


\section{Table of System parameters}
\label{app:planet_data}
All of the stellar and planetary parameters adopted from TEPCat \citep{Tepcat2011} database, for the model simulations of extra 14 exoplanets developed for this work are tabulated in Table \ref{system_para}. The stellar and planetary parameters for the other 20 planets used in this work can be found in \citet{Goyal2020}. First column shows planet names with 'b' omitted indicating first planet of the stellar system as in TEPCat database. Subsequent columns show stellar temperature (T$_{\textup{star}}$) in Kelvin, stellar metallicity ([Fe/H]$_{\textup{star}}$ ), stellar mass (M$_{\textup{star}}$ ) in units of solar mass,  stellar radius (R$_{\textup{star}}$ ) in units of solar radius, logarithmic (base 10) stellar gravity (logg$_{\textup{star}}$ ) in $m/s^2$, semi-major axis (a) in astronomical units, planetary mass (M$_\textup{p}$) in units of Jupiter mass,  planetary radius (R$_\textup{p}$) in units of Jupiter radius, planetary surface gravity (g$_\textup{p}$) in $m/s^2$, planetary equilibrium temperature (Teq$_\textup{p}$) in Kelvin assuming 0 albedo and efficient redistribution, V magnitude (V$_{\textup{mag}}$) of the host star, discovery paper reference (Discovery Paper) and finally the most updated reference.

\begin{figure*}
\centering
\includegraphics[width=1.0\textwidth]{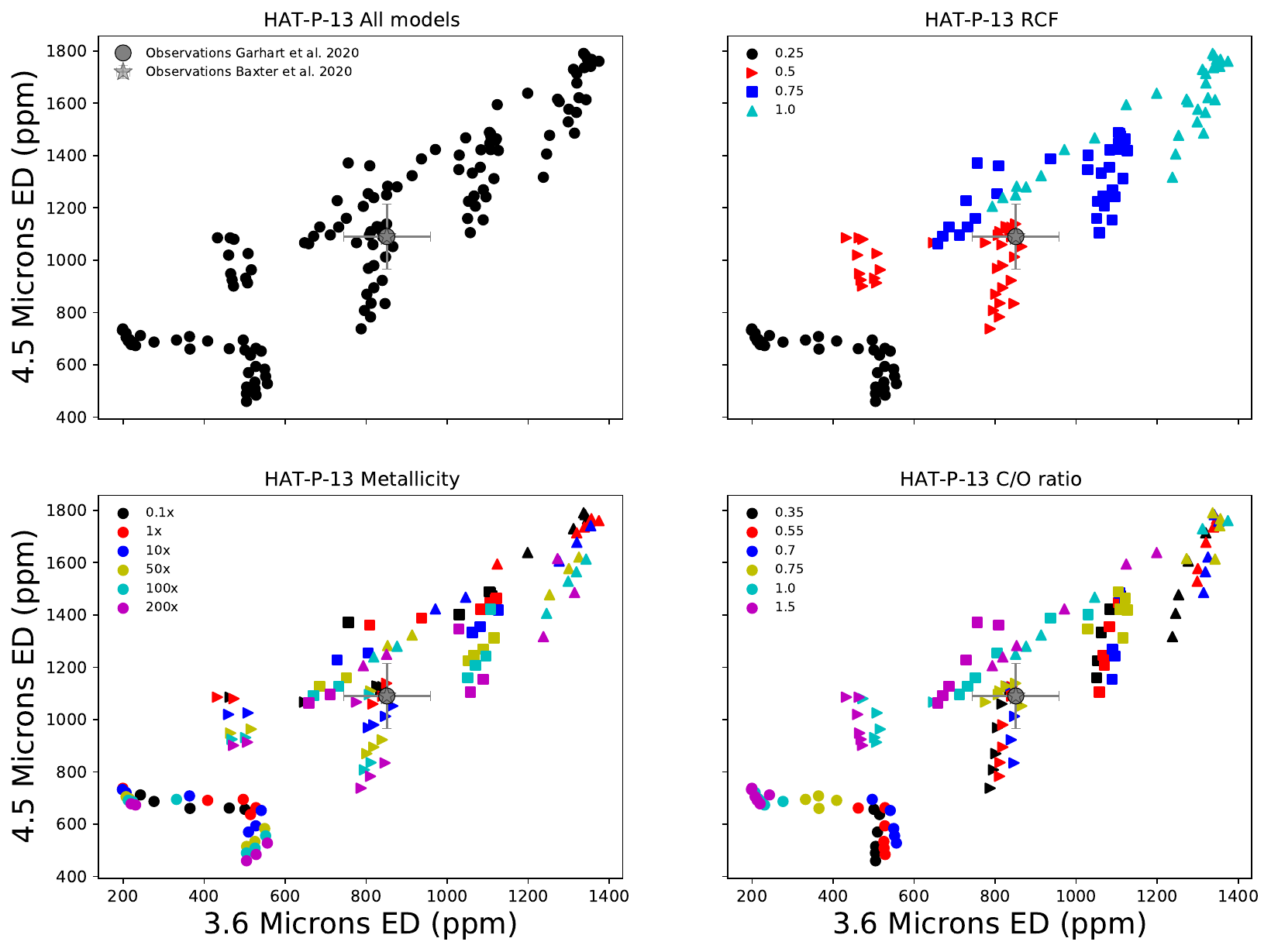}
\caption{The variation of \textit{Spitzer} IRAC channel 1 and 2 model EDs for all models (upper left), the recirculation factors (upper right), metallicity and recirculation factors (lower left), and
C/O ratios and recirculation factors (lower right). The complete figure set for all the planets (34 images) is available in the online journal.}
\label{appfig10}
\end{figure*}

\section{Eclipse Depth Plots}
\label{app:eclipse_plots}
For the reader’s convenience, the figure set of Figure \ref{appfig10} in the online journal version contains four panels showing the variation of \textit{Spitzer} IRAC channel 1 and 2 model EDs for all of the models, models with
different recirculation factors (f$_{\textup{c}}$), models with different f$_{\textup{c}}$ and metallicity, and the models with different f$_{\textup{c}}$ and C/O ratios for all the planets. The \textit{Spitzer} observed EDs in these channels from \citet{Garhart2020} and \citet{Baxter2020} are also shown for each planet in each of the panels. An example plot for HAT-P-13b is shown in Figure \ref{appfig10} here.

\movetabledown=50mm
\begin{rotatetable*}
\begin{deluxetable}{cccccccccccccc}
\tablecaption{System Parameters}

\tabletypesize{\footnotesize} 
\tablecolumns{14}
\tablewidth{15pt}

 \tablehead{
 \colhead{System} &  \colhead{T$_{\textup{star}}$} &	 \colhead{[Fe/H]$_{\textup{star}}$ } &  \colhead{M$_{\textup{star}}$} &  \colhead{R$_{\textup{star}}$} &  \colhead{logg$_{\textup{star}}$}	&  \colhead{a} &  \colhead{M$_\textup{p}$} &  \colhead{R$_\textup{p}$} &  \colhead{g$_\textup{p}$} &  \colhead{Teq$_\textup{p}$} &  \colhead{V$_{\textup{mag}}$} & \colhead{Discovery Paper} & \colhead{Updated Reference} \\
 \colhead{} & \colhead{(K)} & \colhead{} & \colhead{(M$_{\textup{sun}}$)} & \colhead{(R$_{\textup{sun}}$)} & \colhead{($m/s^2$)} & \colhead{(AU)} & \colhead{(M$_{\textup{jup}}$)} & \colhead{(R$_{\textup{jup}}$)} &\colhead{($m/s^2$)} & \colhead{(K)} & \colhead{} & \colhead{} & \colhead{} \\
}

\startdata
KELT-02	&	﻿6151	&	0.034	&	1.314	&	1.836	&	4.03	&	0.05504	&	1.524	&	1.29	&	22.7	&	1712	&	﻿8.7	&	\citet{Beatty2012}	&	\citet{Beatty2012}	\\
KELT-03	&	6306	&	0.044	&	1.278	&	1.472	&	4.209	&	0.04122	&	1.477	&	1.345	&	20.2	&	1811	&	9.82	&	\citet{Pepper2013}	&	\citet{Pepper2013}	\\
Qatar-1	&	4910	&	0.2	&	0.838	&	0.803	&	4.552	&	0.02332	&	1.294	&	1.143	&	24.55	&	1418	&	12.84	&	\citet{Alsubai2011}	&	\citet{Collins2017}	\\
WASP-014	&	6462	&	0	&	1.3	&	1.318	&	4.312	&	0.037	&	7.59	&	1.24	&	123	&	1872	&	9.75	&	\citet{Joshi2009}	&	\citet{Raetz2015}	\\
WASP-018	&	6400	&	0.1	&	1.295	&	1.255	&	4.353	&	0.02055	&	10.52	&	1.204	&	179.9	&	2411	&	9.27	&	\citet{Hellier2009}	&	\citet{Maxted2013}	\\
WASP-036	&	5928	&	-0.01	&	1.081	&	0.985	&	4.486	&	0.02677	&	2.361	&	1.327	&	33.2	&	1733	&	12.7	&	\citet{Smith2012}	&	\citet{Mancini2016}	\\
WASP-046	&	5600	&	-0.37	&	0.828	&	0.858	&	4.489	&	0.02335	&	1.91	&	1.174	&	34.3	&	1636	&	12.93	&	\citet{Anderson2012}	&	\citet{Ciceri2016}	\\
WASP-064	&	5550	&	-0.08	&	1.004	&	1.058	&	4.392	&	0.02648	&	1.271	&	1.271	&	19.6	&	1689	&	12.29	&	\citet{Gillon2013}	&	\citet{Gillon2013}	\\
WASP-065	&	5600	&	-0.07	&	0.93	&	1.01	&	4.4	&	0.0334	&	1.55	&	1.112	&	28.7	&	1480	&	11.9	&	\citet{Chew2013}	&	\citet{Chew2013}	\\
WASP-075	&	6100	&	0.07	&	1.14	&	1.26	&	4.29	&	0.0375	&	1.07	&	1.27	&	15.1	&	1710	&	11.45	&	\citet{Chew2013}	&	\citet{Chew2013}	\\
WASP-077	&	5605	&	0.07	&	1.002	&	0.955	&	4.33	&	0.024	&	1.76	&	1.21	&	27.61	&	1677	&	10.3	&	\citet{Maxted2013}	&	\citet{Maxted2013}	\\
WASP-087	&	6450	&	-0.41	&	1.204	&	1.627	&	4.096	&	0.02946	&	2.18	&	1.385	&	26.1	&	2322	&	10.74	&	\citet{Anderson2014}	&	\citet{Anderson2014}	\\
WASP-100	&	6900	&	-0.03	&	1.57	&	2	&	4.04	&	0.0457	&	2.03	&	1.69	&	16.2	&	2190	&	10.8	&	\citet{Hellier2014}	&	\citet{Hellier2014}	\\
WASP-104	&	5450	&	0.32	&	1.076	&	0.963	&	4.503	&	0.02918	&	1.272	&	1.137	&	22.5	&	1516	&	11.79	&	\citet{Smith2014}	&	\citet{Smith2014}	\\
\enddata
\label{system_para}
\end{deluxetable}
\end{rotatetable*}


\newpage

\bibliography{main}{}
\bibliographystyle{aasjournal}



\end{document}